\documentclass{aa}  
\let\linenumbers\relax

\usepackage{graphicx}
\usepackage{txfonts}
\usepackage{lipsum}
\usepackage{subcaption}        
\usepackage{lscape}            
\usepackage{placeins}           
\usepackage{amsmath}
\usepackage{amssymb}  
\usepackage{subcaption}                                     
\usepackage{multirow}                                       
\usepackage{tablefootnote}

\usepackage{natbib}                                       
\usepackage{aas_macros}                                     
\bibpunct{(}{)}{;}{a}{}{,}                                  

\usepackage{hyperref}                                       
\hypersetup{
    colorlinks=true, 
    linkcolor=blue, 
    urlcolor=blue, 
    citecolor=blue
}

\newcommand{\NcatTotal}{15\,292}        
\newcommand{\NafterCuts}{14\,036}       
\newcommand{\NfitFinal}{14\,017}        
\newcommand{\Nclipped}{19}              
\newcommand{\NpassOne}{12\,468}         
\newcommand{\Nnegplx}{104}              
\newcommand{\Vperpthr}{23}              
\newcommand{\Noutliers}{1474}           
\newcommand{\Foutliers}{10.5}           
\newcommand{\Nadaptive}{1662}           
\newcommand{\Fadaptive}{11.9}           

\newcommand{\Rsun}{8.178}               
\newcommand{\Zsun}{20}                  
\newcommand{\clipthr}{40}               
\newcommand{\polydeg}{4}                
\newcommand{\Nboot}{1000}               
\newcommand{\Chebmin}{4}                
\newcommand{\Chebmax}{13}               
\newcommand{\Uval}{8.06}   \newcommand{\Uerr}{0.41}
\newcommand{\Vval}{10.91}  \newcommand{\Verr}{0.86}
\newcommand{\Wval}{7.53}   \newcommand{\Werr}{0.14}
\newcommand{\Uerrform}{0.150}  \newcommand{\Verrform}{0.245}
\newcommand{\Werrform}{0.114}
\newcommand{\Vcsun}{236.3}  \newcommand{\Vcsunerr}{1.1}
\newcommand{\Oorta}{16.35}  \newcommand{\Oortaerr}{0.21}
\newcommand{\Oortb}{-12.54} \newcommand{\Oortberr}{0.19}
\newcommand{\dVdR}{-3.81}   \newcommand{\dVdRerr}{0.38}
\newcommand{\Omegazero}{28.89}  \newcommand{\Omegazeroerr}{0.13}
\newcommand{\dVcnorun}{1.7}             
\newcommand{\signorun}{1.8}             
\newcommand{\corrVsunVc}{-0.51}         
\newcommand{\corrUsunVc}{-0.11}         

\newcommand{\Rcovmin}{2}                
\newcommand{\Rcovmax}{15}               
\newcommand{\Rsigmin}{5.5}                
\newcommand{\Rsigmax}{11}               
\newcommand{\Fsigrange}{93}             

\newcommand{\sigpmlAll}{2.38}   \newcommand{\sigpmbAll}{1.44}
\newcommand{\sigpmlClean}{1.61} \newcommand{\sigpmbClean}{0.97}
\newcommand{\redpml}{32.5}      \newcommand{\redpmb}{32.4}

\newcommand{\slopeout}{-2.42}  \newcommand{\slopeouterr}{0.59}  
\newcommand{\slopetwo}{-2.47}  \newcommand{\slopetwoerr}{0.25}  

\newcommand{\kurtl}{54}        \newcommand{\kurtb}{90}
\newcommand{\kurtlcln}{39}     \newcommand{\kurtbcln}{26}
\newcommand{\slopeZ}{-1.49}    \newcommand{\slopeZerr}{0.08}
\newcommand{\slopeZin}{-1.81}  \newcommand{\slopeZinerr}{0.10}
\newcommand{\slopeRl}{-0.032}  \newcommand{\slopeRlerr}{0.013}
\newcommand{\slopeRb}{+0.013}  \newcommand{\slopeRberr}{0.008}

\begin{document}

   \title{The Galactic rotation curve from the ALS III catalogue of OB Stars}

   \author{S. R. Berlanas\inst{1,2,3}\fnmsep\thanks{srberlan@iac.es}
        \and M. Pantaleoni González\inst{4}
        \and J. Maíz Apellániz\inst{1}
        }

   \institute{Centro de Astrobiología (CSIC–INTA), Camino Bajo del Castillo,
s/n, E-28692 Villanueva de la Cañada, Madrid, Spain
   \and Instituto de Astrofísica de Canarias, c/ Vía Láctea, s/n, E-38205 La Laguna, Tenerife, Spain
   \and Departamento de Astrofísica, Universidad de La Laguna, E-38206 La Laguna, Tenerife, Spain
   \and University of Vienna, Department of Astrophysics, Türkenschanzstrasse 17, 1180 Wien, Austria}

   \date{Received XX Oct 2026; accepted XX XXX 2026}


  \abstract
   {The Galactic rotation curve $V_{\rm c}(R)$ reveals the radial mass distribution of the Milky Way, but its determination depends on the tracer population employed. Samples of different ages have different kinematics, and the resulting curves differ accordingly. Young massive stars are the least dynamically evolved tracer available, retaining the velocity field of the interstellar medium from which they formed, yet existing determinations from OB star samples are confined to the solar neighbourhood and constrain only the local gradient.}
   {We aim to determine the Galactic rotation curve and the solar motion from a large, spectroscopically confirmed sample of OB stars, using the five-dimensional astrometry from \textit{Gaia} DR3. We characterise the shape of the curve across the disc and assess the limitations of a solution based on tangential data alone.}
   {We use \NfitFinal\ OB stars selected from the ALS III catalogue with revised distances and corrected proper motions. The rotation curve is parameterised as a polynomial of degree \polydeg\ in Galactocentric radius, expressed in the Chebyshev basis, and fitted to the observed proper motions by weighted least squares, solving simultaneously for the solar motion. A two-pass iterative scheme excludes stars closer than $1$\,kpc from the first pass, motivated by the $1/r$ amplification of velocity uncertainties, and reinstates them in the second. Confidence intervals are obtained from \Nboot\ bootstrap resamplings.}
   {We obtain $V_{\rm c}(R_\odot = \Rsun\,{\rm kpc}) = \Vcsun \pm \Vcsunerr$\,km\,s$^{-1}$, with $U_\odot = \Uval \pm \Uerr$\,km\,s$^{-1}$, $V_\odot = \Vval \pm \Verr$\,km\,s$^{-1}$, and $W_\odot = \Wval \pm \Werr$\,km\,s$^{-1}$, and Oort constants $A = \Oorta \pm \Oortaerr$\,km\,s$^{-1}$\,kpc$^{-1}$ and $B = \Oortb \pm \Oortberr$\,km\,s$^{-1}$\,kpc$^{-1}$. The curve rises within the solar circle and declines gently beyond it, with a gradient at the solar radius of ${\rm d}V_{\rm c}/{\rm d}R = \dVdR \pm \dVdRerr$\,km\,s$^{-1}$\,kpc$^{-1}$. The curve is constrained over $\Rsigmin$--$\Rsigmax$\,kpc, the interval containing $\Fsigrange\%$ of the sample. Residuals show no trend with Galactocentric radius, but a trend with height above the plane is present in the latitudinal component.}
   {Young massive stars yield a rotation curve from five-dimensional \textit{Gaia} astrometry alone, demonstrating that spectroscopic follow-up is not a prerequisite for this measurement. Because the tracer is dynamically cold and the curve is obtained by fitting the proper motions directly, the result requires no asymmetric drift correction and does not rest on the assumptions of Jeans modelling, so it is subject to a different set of systematics from determinations based on older tracers. The gentle decline we find beyond the solar radius agrees with those determinations where the two overlap, which makes it unlikely to be an artefact of either approach. The one parameter that tangential data cannot constrain on its own is the solar radial motion, whose formal error understates its true uncertainty. The rotation curve itself is nevertheless unaffected. The stars departing most strongly from the curve, between \Noutliers\ and \Nadaptive\ depending on the selection criterion, constitute the largest sample of OB runaway candidates compiled to date. Their individual identification is deferred to a forthcoming paper based on \textit{Gaia} DR4 (Paper~II). The model is axisymmetric and does not account for spiral-arm perturbations or other non-circular motions, which are left for that analysis.}

  
   \maketitle

\section{Introduction}
\label{sec:intro}

The Galactic rotation curve $V_{\rm c}(R)$ is a fundamental descriptor of the dynamics of the Milky Way, encoding the radial distribution of its mass and providing the boundary conditions for any model of the disc and of the dark matter halo. Its determination is, however, tracer dependent. Stellar populations of different ages have different velocity dispersions and different asymmetric drifts, and a curve derived from one population is not directly interchangeable with a curve derived from another. The choice of tracer is therefore part of the measurement, not incidental to it.

Early determinations relied on the terminal-velocity method applied to \ion{H}{i} and CO line surveys, later supplemented by VLBI astrometry of masers \citep{Reid2014} and, more recently, by \textit{Gaia} \citep[see][for a compilation spanning all of these]{Sofue2009, Sofue2020}. The outer disc, in particular, remains poorly constrained. Usable tracers become scarce beyond the solar vicinity, and it is precisely there that the shape of the curve carries most information about the mass distribution.

Young massive stars offer a distinct advantage in this respect. Their main-sequence lifetimes are between $\sim$3\,Ma and $\sim$30\,Ma, so they remain, with few exceptions, close to their birthplaces and have undergone minimal dynamical heating. They therefore retain, to a good approximation, the velocity field of the interstellar medium out of which they formed, and constitute the dynamically coldest tracer available. Their high intrinsic luminosity makes them detectable across a large fraction of the disc and yields highly precise \textit{Gaia} astrometry. Despite these properties, massive OB stars\footnote{Throughout, OB star denotes a massive star (M > 8 M$_\odot$) of spectral type O or B. In practice, subtype B2 or earlier for dwarfs, B5 or earlier for giants, and B9 or earlier for supergiants \citep{Morgan1951, Maizetal26a}.} have not previously been used to map the rotation curve across a substantial fraction of the disc. 
Existing determinations from OB star samples span baselines of $4$\,kpc or less
\citep{BobylevBajkova2018, BobylevBajkova2019, Almannaei2024, MartinezGarcia2026}, and constrain the local gradient rather than the run of the curve with radius. Samples reaching further have been selected by effective temperature from \textit{Gaia} photometry alone \citep{GaiaDrimmel2023}, a selection that is not clean for OB stars, and were used to map the velocity field rather than to fit a rotation curve.

The ALS III catalogue \citep{PantaleoniGonzalez2025} makes such a measurement possible. It is the most complete compilation of Galactic OB stars with reliable spectroscopic identifications available, and it has been cross-matched with \textit{Gaia} DR3 and provided with distances and proper motions specifically revised for bright stars. Spectroscopic confirmation removes the contamination by evolved or misclassified objects that affects photometrically selected samples. Its span in heliocentric distance is unprecedented for such a sample, reaching well beyond the solar vicinity, so the rotation curve can be mapped as a function of radius over a large fraction of the disc rather than characterised by a local gradient alone. Proper motions and distances suffice to constrain the model, so the measurement can be made from the five-dimensional \textit{Gaia} astrometry alone, without recourse to spectroscopic radial velocities.

The same sample also contains information of a different kind. Not every star follows the fitted curve. A small fraction depart from it by far more than the velocity dispersion of the disc, and we quantify this departure through the peculiar tangential velocity $V_{\perp}$, the residual of the converged model converted into a velocity at the distance of each star (Sect.~\ref{sec:vcobs}). These stars are OB runaway candidates. We count them here with two criteria (Sect.~\ref{sec:robustness}), but their individual identification, characterisation and physical interpretation are deferred to a forthcoming paper \citep[][hereafter Paper~II]{PaperII}, which will make use of \textit{Gaia} DR4. In this paper they enter as a possible source of contamination of the fit, and we show in Sect.~\ref{sec:robustness} that they do not affect it.

The model assumes an axisymmetric velocity field, so non-circular motions, such as the perturbations induced by the spiral arms, are not modelled here. 
Their analysis requires a larger and more precise sample, and we leave it, together with the refined identification of the runaway candidates, for Paper~II.

This paper is organised as follows. Section~\ref{sec:data} describes the ALS III sample, the astrometric corrections and the selection criteria. Section~\ref{sec:methodology} presents the kinematic model, the fitting procedure and the estimation of uncertainties. Section~\ref{sec:results} gives the derived parameters together with the validation tests. Section~\ref{sec:discussion} discusses the gradient of the curve in the outer disc, the determination of the solar radial motion, the influence of local structure, the robustness of the curve against kinematic outliers, and the comparison with previous determinations. Our conclusions are summarised in Section~\ref{sec:conclusions}.

\section{Data and astrometric corrections}
\label{sec:data}

\subsection{The ALS III sample}
\label{sec:sample}

The ALS III catalogue \citep{PantaleoniGonzalez2025} is the most complete
compilation of bright Galactic massive stars published to date. We use the Galactic Luminous Star (GLS) subsample, which comprises objects with reliable, validated spectroscopic identifications as OB stars together with confirmed Wolf--Rayet stars, amounting to \NcatTotal\ objects with \textit{Gaia} DR3 astrometry. Because membership rests on spectroscopy rather than on photometric criteria, the contamination by evolved stars, subdwarfs and misclassified objects that would otherwise bias the kinematics is absent by construction.

The stars are young, with main-sequence lifetimes between $\sim$3\,Ma and $\sim$30\,Ma, and their velocity dispersion is correspondingly small. Their high intrinsic luminosity makes them detectable across a large fraction of the disc, and yields to high-precision \textit{Gaia} astrometry.

\begin{figure*}[t]
\centering
\includegraphics[width=\textwidth]{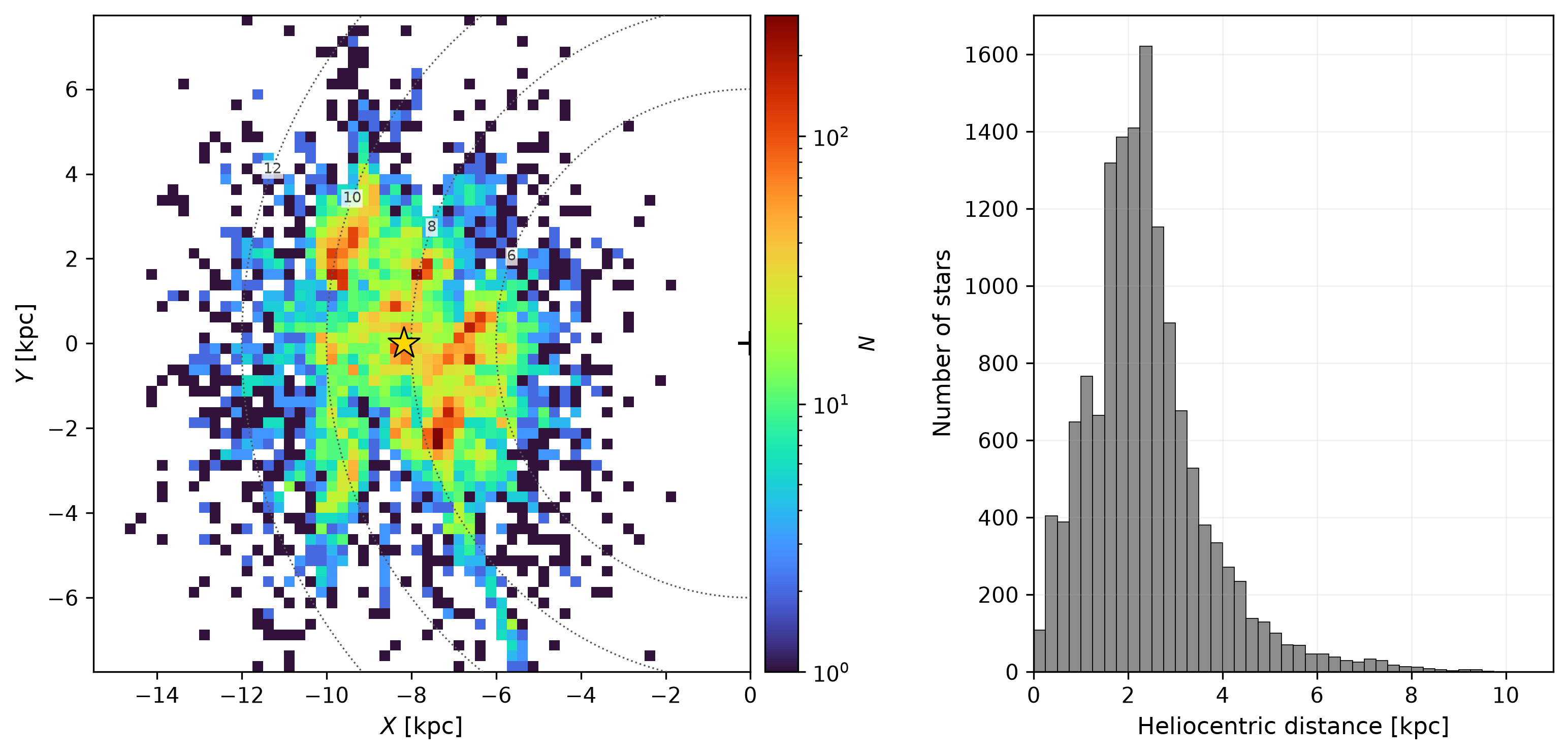}
\caption{Distribution of the \NfitFinal\ ALS III stars used in the fit.
\textit{Left:} projection onto the Galactic plane in Galactocentric Cartesian coordinates, with the Sun at $(X, Y) = (\Rsun, 0)$\,kpc and the Galactic centre at the origin. \textit{Right:} distribution of heliocentric distances.}
\label{fig:sample}
\end{figure*}

The sample spans heliocentric distances up to $\sim 10$\,kpc and covers the full range of Galactic longitude (Fig.~\ref{fig:sample}, which shows the \NfitFinal\ stars retained after the cuts of Sect.~\ref{sec:selection}). The coverage is not uniform, being over-represented towards the nearest spiral arms and towards regions of low extinction, a selection effect we return to in Sect.~\ref{sec:local}.

The observables used in this work are the Galactic coordinates $l$ and $b$, the
distance, and the proper motions, which we transform from equatorial to Galactic components, $\mu_l$ and $\mu_b$. The analysis uses only five-dimensional astrometry. No spectroscopic radial velocities enter the fit, and none are required, although working with purely tangential information has consequences that we examine in Sect.~\ref{sec:degeneracy}.

\subsection{Distances}
\label{sec:distances}

\textit{Gaia} DR3 parallaxes require systematic correction, which is particularly relevant for the bright stars that dominate our sample. Direct use of the published values leads to overconfident parallax uncertainties and to systematically overestimated distances in this magnitude regime. Following \citet{MaizApellaniz2022} a zero-point offset is applied,
\begin{equation}
\varpi_{\rm c} = \varpi - \varpi_{0} ,
\label{eq:plx_zp}
\end{equation}
and the catalogue uncertainty is replaced by an external estimate,
\begin{equation}
\sigma_{\varpi, \mathrm{ext}} = \sqrt{\left(k \, \sigma_{\varpi, \mathrm{int}}\right)^{2} + \sigma_{\varpi, \mathrm{sys}}^{2}} ,
\label{eq:plx_err}
\end{equation}
where $\sigma_{\varpi, \mathrm{int}}$ is the internal \textit{Gaia} uncertainty, $k$ is a scaling factor from the calibrations of \citet{MaizApellaniz2022}, and $\sigma_{\varpi, \mathrm{sys}} = 10.3\,\mu$as is the systematic uncertainty for bright stars \citep{MaizApellaniz2021}. The published DR3 uncertainties are significantly underestimated in this magnitude regime, and using them directly would produce overconfident distances and, through the weighting of the fit, biased kinematic parameters.

Distances are not obtained by direct inversion of the corrected parallax, which would be biased in the sense of \citet{LutzKelker1973}. They are instead derived by \citet{PantaleoniGonzalez2025} with a Bayesian estimator using a prior appropriate for the population under study: a self-gravitating isothermal thin disc with a scale height typical of the OB population, plus a halo component to accommodate objects at large heights above the plane \citep{MaizApellaniz2001}. This is preferable here to a general-purpose prior such as that of \citet{BailerJones2021}, which is built from the global \textit{Gaia} sample and is therefore dominated by low-mass stars with a very
different spatial distribution. 

\subsection{Proper motion corrections}
\label{sec:pm_corr}

Proper motion uncertainties are treated analogously, following
\citet{CantatGaudin2021},
\begin{equation}
\sigma_{\mu, \mathrm{ext}} = \sqrt{\left(k \, \sigma_{\mu, \mathrm{int}}\right)^{2} + \sigma_{\mu, \mathrm{sys}}^{2}} ,
\label{eq:pm_err}
\end{equation}
with a systematic uncertainty $\sigma_{\mu, \mathrm{sys}} = 23\,\mu$as\,yr$^{-1}$ \citep{Lindegren2021}. The correction is applied in three steps. The catalogue uncertainties in $\mu_{\alpha*}$ and $\mu_{\delta}$ are first scaled by the factor $k$; the proper motions are then transformed to Galactic components using the full astrometric covariance matrix of each star, following \citet{ButkevichLindegren2022}; and the systematic term is added in quadrature last. 

Performing the transformation with the complete covariance matrix rather than with the variances alone is essential. \textit{Gaia} DR3 reports ten correlations among the five astrometric parameters, and the rotation between the equatorial and Galactic frames is position dependent, so neglecting them would bias the transformed uncertainties in a way that is itself a function of $l$ and $b$.

\subsection{Sample selection}
\label{sec:selection}

From the full ALS III GLS subsample we retain stars with a relative parallax uncertainty $\sigma_{\varpi}/\varpi < 0.20$, which ensures that the distances are well determined and that the weights entering the fit are meaningful. We further require $\varpi > 0$. This is not implied by the previous criterion, since a negative parallax yields a negative ratio and would formally satisfy it.

Stars with $\varpi < 0$ carry no distance information of their own, in the sense that any distance assigned to them is set by the prior rather than by the measurement, and cannot therefore constrain the velocity field. Most of them (about $70\%$) are assigned distances that place them at $R > \Rcovmax$\,kpc, beyond the range over which the sample provides genuine coverage. A visual inspection indicates that they are not simply distant stars with noisy astrometry but objects with poor astrometric solutions. Retaining them would extend the radial baseline of the analysis into a regime where the distances are set by the prior, in the outer region where the polynomial is least constrained. This cut removes \Nnegplx\ stars.

After these cuts the working sample contains \NafterCuts\ stars. A further selection is applied during the fit itself and is described in Sect.~\ref{sec:fitting}.

\section{Kinematic model and fitting methodology}
\label{sec:methodology}

\subsection{Kinematic model}
\label{sec:model}

 We assume that the stars move on circular orbits about the Galactic centre with a speed $V_{\rm c}(R)$ that depends only on the Galactocentric cylindrical radius $R$, and that the Sun is displaced from that circular motion by the peculiar velocity $(U_\odot, V_\odot, W_\odot)$, defined positive towards the Galactic centre, in the direction of rotation, and towards the North Galactic Pole, respectively. The model is the one developed by \citet{PantaleoniGonzalez2024}, adapted here to a continuous parameterisation of the rotation curve. 

\subsubsection{Solar height and the tilt of the Galactic equator}
\label{sec:tilt}

 We adopt $R_\odot = \Rsun$\,kpc, the distance to Sgr\,A$^{*}$ from \citet{Gravity2019}, and a solar height above the Galactic plane $Z_\odot = \Zsun$\,pc \citep{Maizetal08a}, both held fixed.

The consequence of a non-zero $Z_\odot$ is geometric rather than positional. The Galactic coordinate system is defined so that its origin points at the Galactic centre. If the Sun lies above the mid-plane, the line of sight to the centre is inclined with respect to that plane, and the Galactic equator $b = 0$ cannot be parallel to it. The two planes are tilted with respect to each other by $\arcsin(Z_\odot/R_\odot) \approx 8\farcm4$, and the assumed velocity field must be rotated accordingly before being projected onto the observables. Displacing the stars vertically would have no effect on its own, since a cylindrically symmetric rotation curve is by construction independent of height. It is the tilt that enters the model.

Most published determinations of the rotation curve neglect this tilt. \citet{GaiaDrimmel2023}, for instance, note that the offset of Sgr\,A$^{*}$ from $b = 0$ implies a residual tilt of $\approx 0\fdg1$ and set $Z_\odot = 0$ on the grounds that it is negligible for mapping in the $(X, Y)$ plane. That is reasonable for their purpose. It is not obviously so for ours, since the tilt terms enter the predicted proper motions directly, modifying them at the level of hundredths of mas\,yr$^{-1}$ and corresponding to $\sim 1$\,km\,s$^{-1}$ for stars beyond $6$\,kpc \citep{PantaleoniGonzalez2024}. We therefore retain them throughout, while verifying in Sect.~\ref{sec:model_params} that the derived parameters are insensitive to this choice.

Two quantities follow. The radius of the solar circle, that is, the Galactocentric distance of the Sun projected onto the Galactic plane, is
\begin{equation}
R_{\rm c} = \sqrt{R_\odot^{2} - Z_\odot^{2}} ,
\label{eq:Rc}
\end{equation}
and we write $s \equiv Z_\odot/R_\odot$ for the sine of the tilt angle. Note that $V_{\rm c,\odot} \equiv V_{\rm c}(R_{\rm c})$ is evaluated at $R_{\rm c}$ rather than at $R_\odot$. The two differ by a negligible amount for $Z_\odot = \Zsun$\,pc, but the distinction is kept for consistency.

\subsubsection{Galactocentric radius}

For a star at Galactic coordinates $(l, b)$ and heliocentric distance $r$, the Galactocentric radius follows from projecting its position onto the Galactic plane and applying the cosine rule to the triangle formed by the Galactic centre, the Sun and the star,
\begin{align}
R^{2} = \; & R_{\rm c}^{2} + r^{2}\cos^{2} b - 2\sqrt{1-s^{2}}\,R_{\rm c}\,r\cos b\cos l \nonumber \\
           & - 2 s R_{\rm c}\, r \sin b + s^{2} r^{2}\left(\sin^{2} b - \cos^{2} b \cos^{2} l\right) \nonumber \\
           & + 2 s \sqrt{1-s^{2}}\, r^{2}\cos b \sin b \cos l .
\label{eq:Rgal}
\end{align}
The terms in $s$ vanish when the Sun is placed in the plane. Setting $s = 0$ recovers the familiar planar expression $R^{2} = R_\odot^{2} + r^{2}\cos^{2}b - 2R_\odot r\cos b\cos l$. The derivation is given in Appendix~\ref{APP:transf}.

\subsubsection{Predicted proper motions and radial velocity}

The model takes three inputs per star, $(l, b, r)$, and predicts the two observable proper motions. Writing $\omega \equiv V_{\rm c}(R)/R$ for the angular velocity at the position of the star,
\begin{align}
\mu_{l}^{\rm pred} = \frac{\kappa}{r}\Big[\, & U_\odot \sin l + \left(\omega R_{\rm c} - V_\odot - V_{\rm c,\odot}\right)\cos l \nonumber \\
  & - \omega r \left(\sqrt{1-s^{2}}\cos b + s \sin b \cos l\right)\Big] , \label{eq:pml}\\[4pt]
\mu_{b}^{\rm pred} = \frac{\kappa}{r}\Big[\, & U_\odot \sin b \cos l + \left(V_\odot + V_{\rm c,\odot} - \omega R_{\rm c}\right)\sin b \sin l \nonumber \\
  & - W_\odot \cos b + \omega\, s\, r \sin l \Big] , \label{eq:pmb}
\end{align}
where $\kappa$ converts km\,s$^{-1}$\,pc$^{-1}$ into mas\,yr$^{-1}$. The terms proportional to $s$ in Eqs.~(\ref{eq:pml}) and (\ref{eq:pmb}) are those introduced by the tilt, which vanish when the Sun is placed in the plane. The $\mu_{b}$ tilt term is of particular interest because $\mu_{b}$ is the component that carries most of the information on $W_\odot$, and in principle on $Z_\odot$ itself \citep{PantaleoniGonzalez2024}.

The same model predicts the line-of-sight velocity of each star,
\begin{equation}
V_{r}^{\rm pred} = -U_\odot \cos b \cos l + \left(\omega R_{\rm c} - V_\odot - V_{\rm c,\odot}\right)\cos b \sin l - W_\odot \sin b .
\label{eq:vr}
\end{equation}
This quantity is not constrained by our data, since no radial velocities enter the fit. It is nevertheless a well-defined prediction of the fitted model, and its role in the determination of $U_\odot$ is discussed in Sect.~\ref{sec:degeneracy}.

\subsubsection{Fixed and free parameters}
\label{sec:model_params}

The solar height $Z_\odot$ is held fixed rather than fitted. Equation~(\ref{eq:pmb}) shows that $\mu_{b}$ is formally sensitive to $Z_\odot$ through the tilt, but the sensitivity is weak and $Z_\odot$ couples to the remaining parameters. Attempts to fit it freely drive it to values of several hundred parsecs \citep{PantaleoniGonzalez2024}, far outside the range allowed by direct determinations \citep{Maizetal08a}, and we therefore adopt the literature value. Repeating the entire analysis with $Z_\odot = 0$ changes $V_{\rm c}(R_\odot)$ by $0.02$\,km\,s$^{-1}$ and leaves the solar motion unchanged, confirming that the fitted parameters do not depend on this choice. Likewise, $R_\odot$ is fixed to the Sgr\,A$^{*}$ distance, which is known far more precisely than our data could determine it. 

The circular velocity at the solar radius is not an independent parameter, being fully determined by the fitted rotation curve through $V_{\rm c,\odot} = V_{\rm c}(R_{\rm c})$. The free parameters are therefore the three components of the solar motion and the $n + 1$ coefficients of the rotation curve, whose parameterisation we describe next.

\subsection{Parameterisation of the rotation curve}
\label{sec:param}

We make no assumption about the functional form of $V_{\rm c}(R)$ beyond smoothness, and parameterise it as a polynomial of degree $n$ in Galactocentric radius, expressed in the Chebyshev basis over the interval
$\Chebmin < R < \Chebmax$\,kpc,
\begin{equation}
V_{\rm c}(R) = \sum_{i=0}^{n} V_{i}\, T_{i}(u) , \qquad u = \frac{2R - (\Chebmax + \Chebmin)}{\Chebmax - \Chebmin} ,
\label{eq:poly}
\end{equation}
where $T_{i}$ is the Chebyshev polynomial of the first kind of order $i$ and $u \in [-1, 1]$ over the adopted interval. All coefficients $V_{i}$ have units of km\,s$^{-1}$. A conventional expansion in powers of $R$ describes the same family of functions but is poorly conditioned over this radial range. Adjacent monomials are nearly indistinguishable over an interval that does not include the origin, so their coefficients become strongly anticorrelated, individual coefficients are recovered with poor precision, and the solution drifts along the resulting flat direction of parameter space until it meets whatever bounds are imposed on it.
We use a Chebyshev basis over this interval for its numerical conditioning. No bounds on the coefficients are required, and each coefficient of the adopted fit is determined to within about $25\%$. The coefficients are nevertheless correlated (the largest correlation is $0.94$), so it is the curve $V_{\rm c}(R)$ as a whole, rather than any single coefficient, that is constrained by the data.

A continuous parameterisation, rather than free values at a set of radii, also addresses a specific limitation of the latter. Sampling $V_{\rm c}(R)$ at $n$ free points and interpolating between them, as done by \citet{PantaleoniGonzalez2024}, couples the value of the curve near the Sun to the solar azimuthal motion $V_\odot$. The two enter the proper motion equations in the combination $\omega R_{\rm c} - V_\odot - V_{\rm c,\odot}$, so a free node placed close to $R_\odot$ is largely degenerate with $V_\odot$. That work reports a correlation coefficient of $-0.97$ between the two, and finds that the fit becomes unphysical as the number of nodes is increased, so that the number of nodes had to be reduced to two. A global polynomial mitigates this coupling, because no single coefficient controls the curve locally at $R_\odot$. In our fit the correlation between $V_\odot$ and $V_{\rm c}(R_\odot)$, measured across the bootstrap replicates, is $\corrVsunVc$. The coupling is reduced but not removed, since it originates in the combination in which the two quantities enter Eqs.~(\ref{eq:pml}) and (\ref{eq:pmb}), and no choice of basis for $V_{\rm c}(R)$ can eliminate it. $V_\odot$ consequently carries the largest bootstrap uncertainty of the three solar motion components (Sect.~\ref{sec:res_solar}).

We adopt $n = \polydeg$, although degrees $3$ to $6$ were tested (Appendix~\ref{APP:degree}). $V_{\rm c}(R_\odot)$ varies by less than $1$\,km\,s$^{-1}$ across them and the local gradient is negative and significant in every case, so the results reported here do not depend on this choice.
Degree $\polydeg$ is not the formally preferred model, since the Bayesian information criterion \citep[BIC;][]{Schwarz1978} marginally favours degree $5$. The BIC is, however, only indicative. The residuals, dominated by intrinsic dispersion and non-circular motions, are not independent, so differences of a few units between neighbouring degrees should not be regarded as decisive. We adopt degree $\polydeg$ because it is the highest degree at which every coefficient remains individually significant, whereas degree $5$ buys its slightly better BIC with a coefficient consistent with zero. Degree $3$ has the least freedom and gives a somewhat shallower local gradient, although consistent within the uncertainties, and is disfavoured by the BIC. 

Polynomials diverge rapidly outside the range constrained by the data. The sample provides coverage over roughly $\Rcovmin$--$\Rcovmax$\,kpc, but the number of stars per $500$\,pc bin falls from $1131$ between $10.0$ and $10.5$\,kpc to $332$ between $10.5$ and $11.0$\,kpc, and from $323$ between $5.5$ and $6.0$\,kpc to $93$ between $5.0$ and $5.5$\,kpc. At both ends the shape of the fit is governed by the extrapolation of the polynomial rather than by the data. We therefore restrict all quantitative statements to $\Rsigmin < R < \Rsigmax$\,kpc, the interval containing $\Fsigrange\%$ of the sample, and display the curve outside it only as an indication.

\subsection{Two-pass iterative fitting}
\label{sec:fitting}

The model is fitted to the observed proper motions by weighted non-linear least
squares, minimising
\begin{equation}
\chi^{2} = \sum_{i} \left[ \left( \frac{\mu_{l,i}^{\rm obs} - \mu_{l,i}^{\rm pred}}{\sigma_{\mu_{l},i}} \right)^{2} + \left( \frac{\mu_{b,i}^{\rm obs} - \mu_{b,i}^{\rm pred}}{\sigma_{\mu_{b},i}} \right)^{2} \right] ,
\label{eq:chi2}
\end{equation}
over the three components of the solar motion and the $n + 1$ polynomial coefficients $V_{0} \ldots V_{n}$, all fitted simultaneously, using the Levenberg--Marquardt algorithm as implemented in \texttt{lmfit} \citep{lmfit}. For $n = \polydeg$ this amounts to eight free parameters. The solar position,
$R_\odot$ and $Z_\odot$, is held fixed.

At each iteration, stars whose total proper motion residual $|\delta\vec{\mu}| = (\delta\mu_{l}^{2} + \delta\mu_{b}^{2})^{1/2}$ exceeds $\clipthr$\,mas\,yr$^{-1}$ are removed and the fit is repeated until neither the retained sample nor the parameters change further.
This is an absolute angular threshold, not a multiple of the residual dispersion, and it is chosen to reject pathological astrometry rather than anomalous kinematics. A residual of $\clipthr$\,mas\,yr$^{-1}$ exceeds anything expected from a disc star at the distances spanned by our sample. At $1$\,kpc it corresponds to a transverse velocity in excess of $180$\,km\,s$^{-1}$, and the implied velocity grows linearly with distance. Stars exceeding it are therefore not fast-moving disc members but objects with a failed astrometric solution, an unresolved companion perturbing the five-parameter fit, or a spurious cross-identification. Consistently with that interpretation, the criterion removes only \Nclipped\ objects out of \NafterCuts.

\subsubsection{First pass: exclusion of nearby stars}
\label{sec:pass1}

In the first pass, run for ten iterations, we exclude stars closer than $1$\,kpc, a limit chosen to enclose the Gould Belt and the nearest OB associations with some margin. The motivation follows from the structure of Eqs.~(\ref{eq:pml}) and (\ref{eq:pmb}), where velocities are converted into angular quantities through a factor $1/r$. Any mismatch between the true and the modelled space velocity of a star, $\delta V$, propagates into a proper motion residual $\delta\mu \propto \delta V / r$. For a peculiar velocity of a few km\,s$^{-1}$ the induced residual is an order of magnitude larger at $r = 0.5$\,kpc than at $r = 5$\,kpc.

These residuals are moreover divided by observational uncertainties that are small for nearby bright stars, so the nearby stars acquire a weight in $\chi^{2}$ out of proportion to the information they carry about the large-scale rotation curve. What they do carry is information about the local velocity field, and that is precisely where the axisymmetric assumption is weakest. Within $1$\,kpc the kinematics are dominated by the Gould Belt and by individual OB associations \citep{Zucker2022}, structures that a smooth polynomial cannot represent. Left in the first pass, these stars drive the solution towards a fit of local structure rather than of Galactic rotation. The solution converges well before the tenth iteration, with neither the retained sample nor the parameters changing after the fifth.

\subsubsection{Second pass: full distance range}
\label{sec:pass2}

Once a stable solution has been obtained from the distant stars, the nearby stars are reinstated and the fit is repeated over the full sample for five further iterations, using the first-pass parameters as the starting point. Convergence is again reached quickly, by the third iteration. This serves two purposes. It refines the parameters using all the available data, and, with a reliable global model in hand, it identifies which stars genuinely depart from the rotation curve, since their residuals are now evaluated against a solution that they have not themselves distorted.

It is important to distinguish this outlier rejection from the identification of runaway stars. The $|\delta\vec{\mu}|$ cut applied here is a robustness device internal to the fitting procedure, targeting pathological astrometry, and it discards only \Nclipped\ objects. The runaway candidates, by contrast, are identified \emph{a posteriori} from the residuals of the converged model through their peculiar tangential velocity $V_{\perp}$ (Eq.~\ref{eq:vperp}), a physical rather than an astrometric quantity, and they are not removed from the fit that produces our nominal rotation curve. Their influence on the result is assessed separately in Sect.~\ref{sec:robustness}, and their individual identification and characterisation are deferred to Paper~II.

\subsection{Peculiar tangential velocity and projected residual velocity}
\label{sec:vcobs}

The fit itself operates on proper motions, but it is useful to project the data onto the $V$--$R$ plane for display. For each star we compute the residual
proper motions
\begin{equation}
\delta\mu_{l} = \mu_{l}^{\rm obs} - \mu_{l}^{\rm pred} , \qquad
\delta\mu_{b} = \mu_{b}^{\rm obs} - \mu_{b}^{\rm pred} ,
\label{eq:resid}
\end{equation}
which measure the departure of each star from the fitted rotation curve. Their combination, converted into a velocity at the distance of the star, gives the peculiar tangential velocity
\begin{equation}
V_{\perp} = r \sqrt{\delta\mu_{l}^{2} + \delta\mu_{b}^{2}} \; \kappa^{-1} ,
\label{eq:vperp}
\end{equation}
that is, the component of the peculiar velocity perpendicular to the line of sight, measured with respect to the local circular motion predicted by our model.

$V_{\perp}$ is the quantity on which the identification of runaway candidates is based. Two criteria are applied. The first is a fixed threshold of $V_{\perp} \geq \Vperpthr$\,km\,s$^{-1}$, the velocity above which a star is more likely than not to be a runaway, obtained by fitting a Maxwell--Boltzmann distribution to the peculiar velocities of the disc population and locating the point at which the observed density exceeds the modelled one by a factor of two \citep{MartinezGarcia2026}. The second, detailed in Paper~II, adapts the threshold to the local dispersion of $V_{\perp}$ across the Galactic plane, recognising that regions of active star formation are kinematically hotter than the smooth disc. Either way, a threshold on $V_{\perp}$ treats nearby and distant stars on the same footing, unlike one on the proper motion residual itself, which is distance dependent because a given peculiar velocity produces an angular residual scaling as $1/r$.

Two limitations apply. The first follows from the absence of radial velocities. $V_{\perp}$ is the projection onto the plane of the sky of a three-dimensional peculiar velocity whose line-of-sight component is unmeasured, and is therefore a lower limit to the true peculiar speed of each star. A star ejected close to the line of sight has a small $V_{\perp}$ regardless of how fast it is moving and will not be recovered, so any sample selected on $V_{\perp}$ is incomplete by construction and the resulting runaway fraction is a lower bound. This incompleteness is quantified in Paper~II. The second is that $V_{\perp}$ is measured with respect to an axisymmetric model, so it also contains the contribution of the large-scale non-circular motions that the model does not describe (Sect.~\ref{sec:local}).

Projecting the residuals instead onto the direction of Galactic rotation and adding the result to the model prediction gives what we call the projected residual velocity of an individual star,
\begin{equation}
V_{\rm res}^{\rm proj} = V_{\rm c}(R) + r \left( \delta\mu_{l} \sin l + \delta\mu_{b} \cos l \right) \kappa^{-1} ,
\label{eq:vcobs}
\end{equation}
where the residuals retain their sign. $V_{\rm res}^{\rm proj}$ is not the azimuthal velocity $V_{\varphi}$ of the star. It is the projection onto the plane of the sky of the whole residual velocity, and which component of a non-circular motion it picks up depends on the Galactic longitude. A peculiar radial velocity is recovered almost entirely near $l \approx 90^{\circ}$, whereas a purely azimuthal one is not recovered at $l \approx 0^{\circ}$ and $180^{\circ}$. Individual values of $V_{\rm res}^{\rm proj}$ are moreover noisy, since they attribute the whole peculiar motion of each star to a single direction. They are shown in Fig.~\ref{fig:rotcurve} for display only, and play no part in the fit, which operates on the proper motions throughout.

\subsection{Uncertainty estimation}
\label{sec:bootstrap_method}

Formal uncertainties derived from the covariance matrix of a least-squares fit assume Gaussian, independent residuals and a correctly specified model. Neither assumption holds here (see Sect.~\ref{sec:residuals}). The residuals contain the imprint of spiral structure and of unmodelled local motions, and are therefore spatially correlated. We consequently estimate confidence intervals by bootstrap resampling, drawing \Nboot\ samples of the same size as the fitting sample with replacement and repeating the full fit for each. This captures the true variability of the fit rather than the propagation of measurement noise alone. The resampling is applied to the converged sample, so the outlier rejection is not repeated within each replicate and the resulting uncertainties do not include the variability of that selection; given that the rejection discards only \Nclipped\ stars, this contribution is negligible.

\begin{figure*}[t]
\centering
\includegraphics[width=\textwidth]{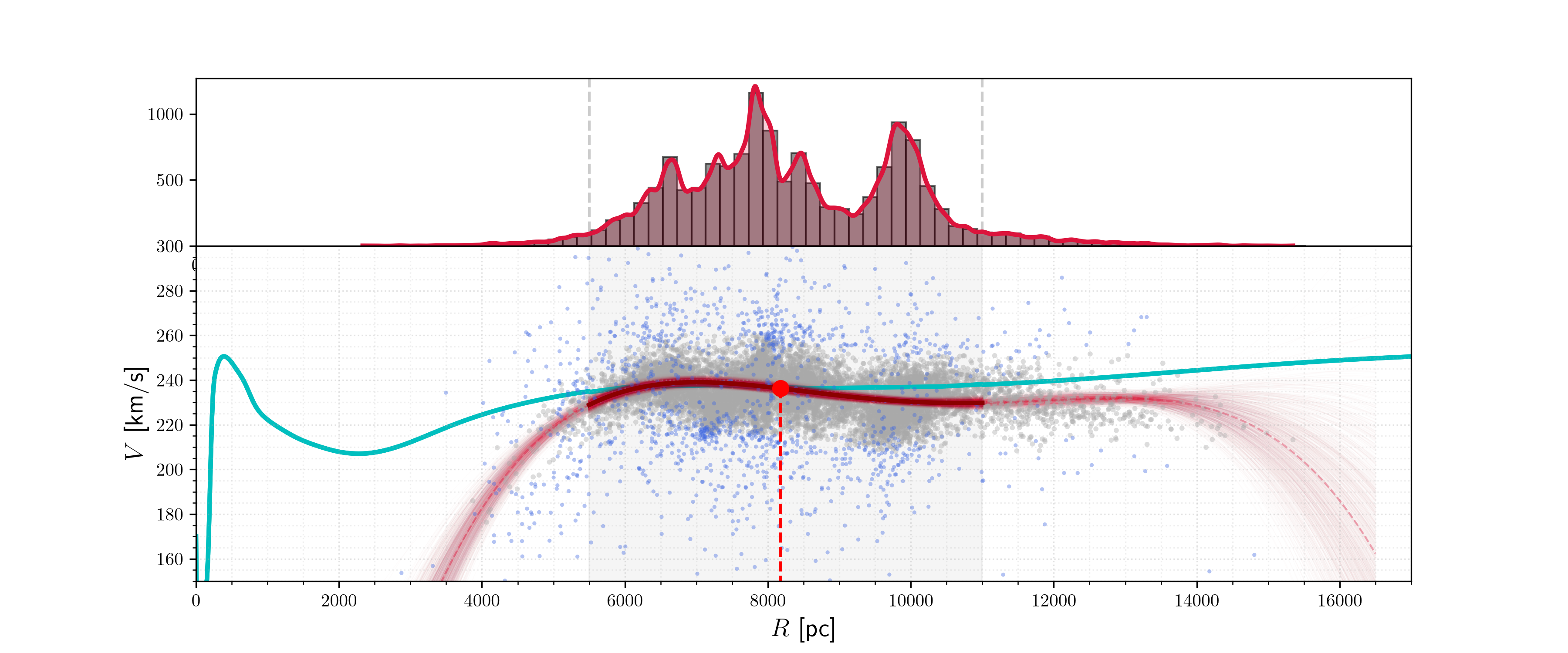}
\caption{Galactic rotation curve derived from the ALS III sample of OB stars used in this work. \textit{Top:} distribution of the fitting sample in Galactocentric radius, with a kernel density estimate overlaid (red). \textit{Bottom:} best-fitting polynomial of degree \polydeg\ (dashed red line) with the family of \Nboot\ bootstrap curves (thin red lines), together with the projected residual velocities $V_{\rm res}^{\rm proj}$ of Eq.~(\ref{eq:vcobs}) for individual stars (grey points). These are not azimuthal velocities $V_{\varphi}$: depending on the Galactic longitude they mix the radial and azimuthal components of any non-circular motion, and they play no part in the fit. The curve and the bootstrap family are drawn solid over the range $\Rsigmin$--$\Rsigmax$\,kpc, where the sample provides statistical weight, and faintly beyond it, where the shape is set by the extrapolation of the polynomial rather than by the data. Stars with a peculiar tangential velocity $V_{\perp} \geq \Vperpthr$\,km\,s$^{-1}$, the runaway candidates selected with the fixed threshold (Sect.~\ref{sec:robustness}), are plotted in blue. The unified rotation curve of \citet{Sofue2020} is shown in cyan for reference. The vertical dashed line and the red dot mark the position of the Sun, and the shaded band the range $\Rsigmin$--$\Rsigmax$\,kpc over which our determination is statistically significant.}
\label{fig:rotcurve}
\end{figure*}

Uncertainties are quoted as the $1\sigma$ dispersion of the bootstrap distribution of each derived quantity, and the $16$th--$84$th percentile envelope of the family of curves is shown in Fig.~\ref{fig:rotcurve}. Derived quantities such as $V_{\rm c}(R_\odot)$ and the Oort constants are recomputed from the coefficients within each bootstrap replicate and their dispersion measured afterwards, rather than propagated analytically from the individual coefficient errors. The coefficients remain correlated even in the Chebyshev basis, and neglecting those correlations would misstate the uncertainty.

\section{Results}
\label{sec:results}

\begin{table}[t]
\centering
\caption{Solar motion parameters derived from the ALS III -- GLS sample of OB stars considered in this work. The second column gives the formal $1\sigma$ error from the least-squares fit. The third column, the $1\sigma$ dispersion of the bootstrap distribution. We adopt the latter as our uncertainty.}
\label{tab:solar_motion}
\begin{tabular}{lrrr}
\hline\hline
Parameter & Value & Formal & Bootstrap \\
\hline
$U_\odot$ (km\,s$^{-1}$) & \Uval  & \Uerrform & \Uerr \\
$V_\odot$ (km\,s$^{-1}$) & \Vval  & \Verrform & \Verr \\
$W_\odot$ (km\,s$^{-1}$) & \Wval  & \Werrform & \Werr \\
\hline
\end{tabular}
\end{table}

\subsection{Solar motion}
\label{sec:res_solar}

Table~\ref{tab:solar_motion} lists the derived solar motion. $W_\odot$ is the best determined of the three and agrees well with determinations based on young populations. $U_\odot$ and $V_\odot$ carry substantially larger uncertainties. $V_\odot$ is affected by its residual coupling to the rotation curve (see Sect.~\ref{sec:param}), and $U_\odot$ is constrained only in combination with the unobserved line-of-sight velocity field, an effect examined in Sect.~\ref{sec:degeneracy}. The bootstrap uncertainties exceed the formal errors throughout, by factors of $1.2$ for $W_\odot$ and $2.7$ and $3.5$ for $U_\odot$ and $V_\odot$, and we adopt the bootstrap values.

\begin{table}[t]
\centering
\caption{Coefficients of the Chebyshev expansion of Eq.~(\ref{eq:poly}), defined over $\Chebmin < R < \Chebmax$\,kpc. All coefficients are in km\,s$^{-1}$. Uncertainties are the $1\sigma$ dispersion of the bootstrap distribution.} 
\label{tab:vc_coeff}
\begin{tabular}{lrr}
\hline\hline
Coefficient & Value & Uncertainty \\
\hline
$V_0$ &  224.62 & 1.31 \\
$V_1$ &   14.03 & 2.23 \\
$V_2$ &  $-13.88$ & 1.53 \\
$V_3$ &   10.61 & 1.10 \\
$V_4$ &  $-3.51$ & 0.89 \\
\hline
$V_{\rm c}(R_\odot)$ & \Vcsun & \Vcsunerr \\
\hline
\end{tabular}
\end{table}

\subsection{The rotation curve}
\label{sec:res_curve}

The fitted coefficients are given in Table~\ref{tab:vc_coeff} and the resulting curve is shown in Fig.~\ref{fig:rotcurve}. Within the solar circle the curve rises with radius, in agreement with previous determinations. Beyond $R \approx 8$\,kpc it declines gently, with a gradient at the solar radius of ${\rm d}V_{\rm c}/{\rm d}R|_{R_\odot} = \dVdR \pm \dVdRerr$\,km\,s$^{-1}$\,kpc$^{-1}$. The implications of this negative gradient are discussed in Sect.~\ref{sec:outer}.
As noted in Sect.~\ref{sec:param}, the decline we report is measured within, not beyond, the significant range.

\subsection{Circular velocity and Oort constants at the solar radius}
\label{sec:res_vc}

Evaluating Eq.~(\ref{eq:poly}) at $R = R_\odot$ gives
\begin{equation}
V_{\rm c}(R_\odot = \Rsun\,\mathrm{kpc}) = \Vcsun \pm \Vcsunerr \; \mathrm{km\,s^{-1}} ,
\label{eq:vcsun}
\end{equation}
a relative precision of $\sim 0.5\%$, where the quoted uncertainty is the bootstrap $1\sigma$ dispersion and therefore already accounts for the correlations among the coefficients. Note that it is a statistical uncertainty and does not include the systematic effect of the non-circular motions that the model does not describe (Sect.~\ref{sec:local}).
The corresponding angular rotation rate is $\Omega_0 = V_{\rm c}(R_\odot)/R_\odot = \Omegazero \pm \Omegazeroerr$\,km\,s$^{-1}$\,kpc$^{-1}$, the quantity most directly comparable with determinations that solve for the angular velocity rather than the linear one.

The Oort constants follow from the value and the local gradient of the rotation curve,
\begin{equation}
A = \frac{1}{2}\left( \frac{V_{\rm c}(R_\odot)}{R_\odot} - \left.\frac{{\rm d}V_{\rm c}}{{\rm d}R}\right|_{R_\odot} \right) , \qquad
B = -\frac{1}{2}\left( \frac{V_{\rm c}(R_\odot)}{R_\odot} + \left.\frac{{\rm d}V_{\rm c}}{{\rm d}R}\right|_{R_\odot} \right) ,
\label{eq:oort}
\end{equation}
where the derivative is obtained from Eq.~(\ref{eq:poly}) by the chain rule,
\begin{equation}
\left.\frac{{\rm d}V_{\rm c}}{{\rm d}R}\right|_{R_\odot} = \frac{2}{\Chebmax - \Chebmin} \sum_{i=1}^{n} V_{i}\, T_{i}'(u_\odot) ,
\label{eq:slope}
\end{equation}
with $u_\odot$ the value of $u$ at $R = R_\odot$. We obtain ${\rm d}V_{\rm c}/{\rm d}R|_{R_\odot} = \dVdR \pm \dVdRerr$\,km\,s$^{-1}$\,kpc$^{-1}$, $A = \Oorta \pm \Oortaerr$ and $B = \Oortb \pm \Oortberr$\,km\,s$^{-1}$\,kpc$^{-1}$, with uncertainties obtained by evaluating Eqs.~(\ref{eq:oort}) and (\ref{eq:slope}) in each bootstrap replicate.

\begin{table}[t]
\centering
\caption{Effect of the two-pass scheme. Parameters and their bootstrap $1\sigma$ uncertainties at the end of the first pass ($r > 1$\,kpc only) and of the second pass (full sample), from \Nboot\ replicates in each case.}
\label{tab:bootstrap}
\begin{tabular}{lrrr}
\hline\hline
Parameter & First pass & Second pass & Shift \\
\hline
$V_{\rm c}(R_\odot)$ (km\,s$^{-1}$) & $239.54 \pm 1.31$ & $236.25 \pm 1.09$ & $-3.29$ \\
$U_\odot$ (km\,s$^{-1}$)            & $6.30 \pm 0.30$   & $8.06 \pm 0.41$   & $+1.75$ \\
$V_\odot$ (km\,s$^{-1}$)            & $8.53 \pm 1.12$   & $10.91 \pm 0.86$  & $+2.37$ \\
$W_\odot$ (km\,s$^{-1}$)            & $7.89 \pm 0.17$   & $7.53 \pm 0.14$   & $-0.36$ \\
\hline
$N$ stars                            & \NpassOne & \NfitFinal & \\
\hline
\end{tabular}
\end{table}

\subsection{Residuals}
\label{sec:residuals}

\begin{figure}[t]
\centering
\includegraphics[width=\columnwidth]{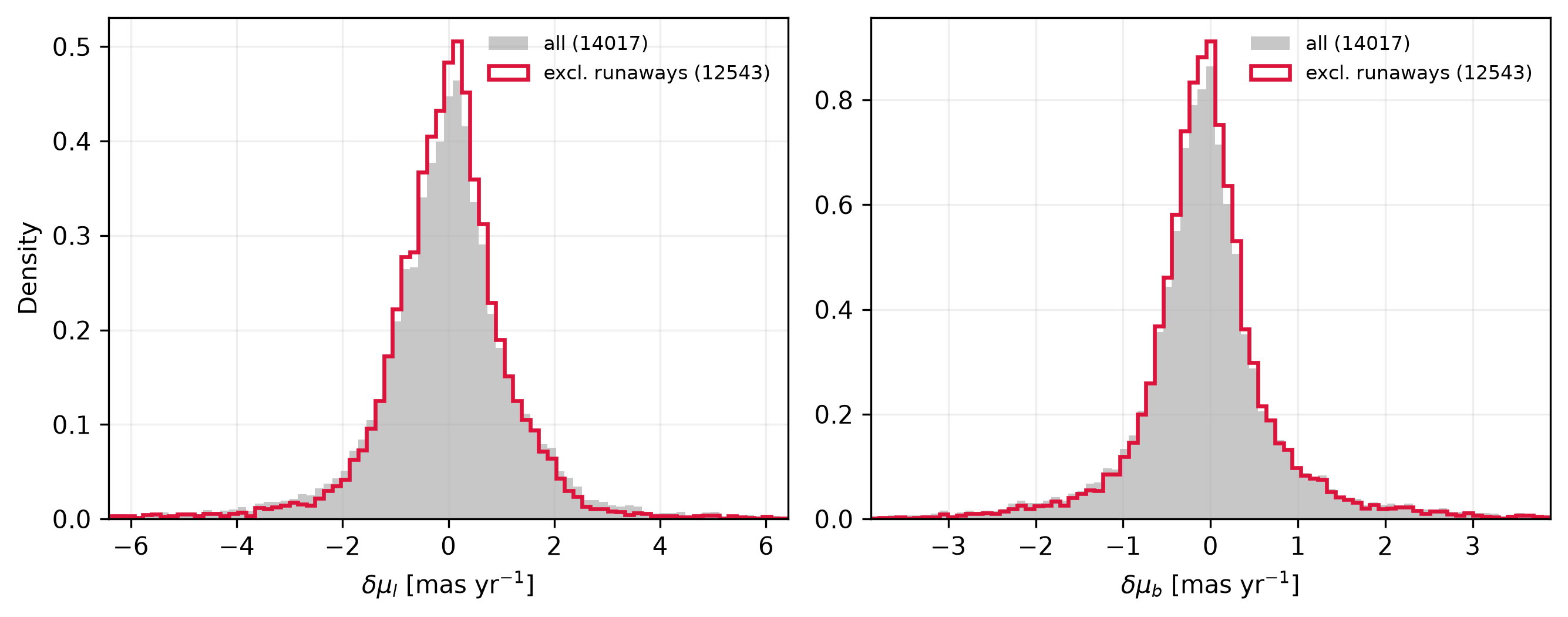}
\caption{Distributions of the proper motion residuals $\delta\mu_{l}$ (left) and $\delta\mu_{b}$ (right) for the full fitting sample (grey) and after removing the kinematic outliers (red). The axes are truncated at four times the dispersion of the latter. }
\label{fig:residual_dist}
\end{figure}

The residuals of the converged fit are shown in Fig.~\ref{fig:residual_dist}. They are centred on zero, with dispersions of $\sigpmlAll$ and $\sigpmbAll$\,mas\,yr$^{-1}$ in $\delta\mu_{l}$ and $\delta\mu_{b}$. Excluding the kinematic outliers these fall to $\sigpmlClean$ and $\sigpmbClean$\,mas\,yr$^{-1}$, reductions of \redpml\% and \redpmb\%. This is expected by construction, since the outliers are selected on the residuals themselves.

Uncertainties derived from the covariance matrix of a least-squares fit assume that the residuals are Gaussian and that each measurement is independent. Neither assumption is satisfied here, which is why we adopt bootstrap uncertainties throughout (see Sect.~\ref{sec:bootstrap_method}).  
Departures from Gaussianity are significant. The excess kurtosis of the residual distributions is $\kurtl$ and $\kurtb$ for $\delta\mu_{l}$ and $\delta\mu_{b}$, against a standard error of $0.04$ at this sample size, and remains at $\kurtlcln$ and $\kurtbcln$ after the outliers are removed. The distributions are therefore strongly peaked with extended tails throughout the sample, not only in the objects we exclude,  as expected of residuals that measure the intrinsic velocity dispersion of the population and the imprint of unmodelled structure rather than observational noise alone. Independence, in turn, is compromised by the clustering of the sample. Stars belonging to the same association or spiral arm share a common peculiar motion, so their residuals carry largely redundant information, whereas the covariance matrix treats all \NfitFinal\ stars as separate measurements. The two effects together make the bootstrap uncertainties exceed the formal errors by factors of $1.2$ to $3.5$.

\begin{figure}[t]
\centering
\includegraphics[width=\columnwidth]{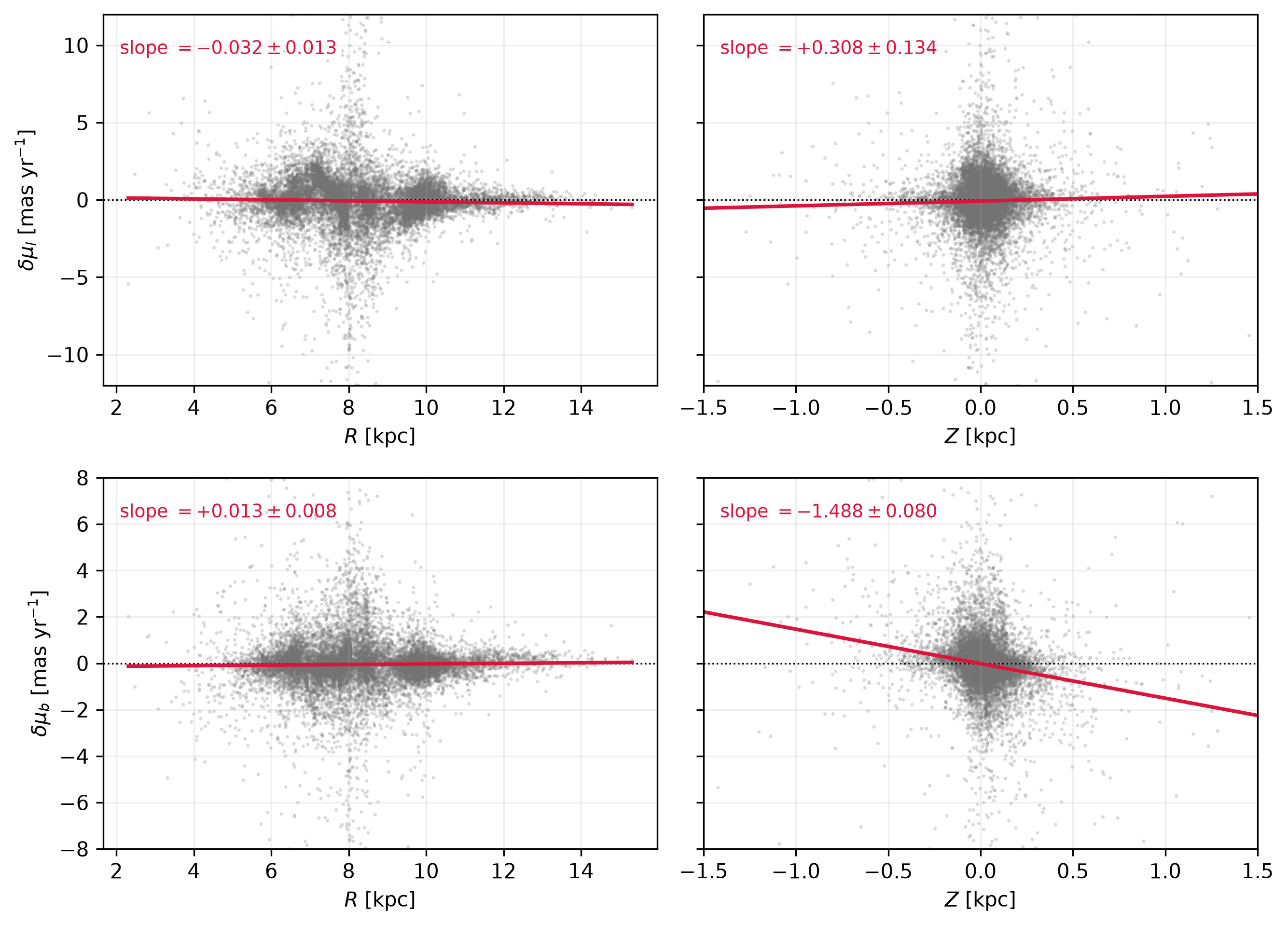}
\caption{Proper motion residuals $\delta\mu_{l}$ (top) and $\delta\mu_{b}$
(bottom) as a function of Galactocentric radius $R$ (left) and of height above the Galactic plane $Z$ (right), for the \NfitFinal\ stars of the fitting sample. The solid line is a linear fit and its slope, in mas\,yr$^{-1}$\,kpc$^{-1}$, is given in each panel. The $Z$ axes are truncated at $\pm 1.5$\,kpc. Only $\delta\mu_{b}$ shows a significant trend, and only with $Z$.}
\label{fig:residual_trends}
\end{figure}

No significant trend is found with Galactocentric radius. The fitted slopes are $\slopeRl \pm \slopeRlerr$ and $\slopeRb \pm \slopeRberr$\,mas\,yr$^{-1}$\,kpc$^{-1}$ for $\delta\mu_{l}$ and $\delta\mu_{b}$ (see Fig.~\ref{fig:residual_trends}), which amount to less than $0.2$\,mas\,yr$^{-1}$ over the radial range of the fit, a small fraction of the dispersion of the residuals. These formal uncertainties assume independent residuals and therefore understate the true ones (Sect.~\ref{sec:bootstrap_method}). However,  a trend with height above the plane is present in $\delta\mu_{b}$, with a slope of $\slopeZ \pm \slopeZerr$\,mas\,yr$^{-1}$\,kpc$^{-1}$ over the full range and $\slopeZin \pm \slopeZinerr$ restricted to $|Z| < 0.5$\,kpc, where $98\%$ of the sample lies.
Because it is much stronger in the latitudinal component, it may point to a vertical motion that the model does not describe. Repeating the entire analysis with $Z_\odot = 0$ changes the slope by less than a tenth of its uncertainty and leaves all other parameters unchanged, so it is not an artifact of the adopted solar height. Its possible origins are discussed in Sect.~\ref{sec:local}.

Two further tests probe how the solution responds to changes in the sample and in the model. Table~\ref{tab:bootstrap} shows the effect of reinstating the nearby stars. $V_{\rm c}(R_\odot)$ decreases by $3.3$\,km\,s$^{-1}$ and the in-plane solar motion shifts by $+1.8$ and $+2.4$\,km\,s$^{-1}$ in $U_\odot$ and $V_\odot$, while $W_\odot$ moves by only $-0.4$\,km\,s$^{-1}$. That the shift is confined to the in-plane components is consistent with the interpretation of Sect.~\ref{sec:pass1}, the nearby sample being dominated by structures whose coherent motion lies in the disc plane. The parameterisation itself is not a factor, since polynomial degrees between $3$ and $6$ yield consistent values of $V_{\rm c}(R_\odot)$ (see Appendix~\ref{APP:degree}).

\section{Discussion}
\label{sec:discussion}
\subsection{The gradient of the rotation curve in the outer disc}
\label{sec:outer}

The most notable feature of the curve derived here is its gentle decline beyond $R \approx 8$\,kpc. We characterise it by two quantities. The local gradient at the solar radius, where the sample is densest, is ${\rm d}V_{\rm c}/{\rm d}R|_{R_\odot} = \dVdR \pm \dVdRerr$\,km\,s$^{-1}$\,kpc$^{-1}$, negative at about $10\sigma$. Averaged over $R_\odot$ to $\Rsigmax$\,kpc the gradient is $\slopeout \pm \slopeouterr$\,km\,s$^{-1}$\,kpc$^{-1}$, and every bootstrap replicate returns a negative value. The two differ because the curve has curvature over this interval, and the second is the quantity directly comparable with determinations that fit a straight line over a long baseline.

Both are recovered at every polynomial degree with the flexibility to follow the shape of the curve. Degrees $4$ to $6$ give mean gradients of $-2.42$, $-2.44$ and $-2.06$\,km\,s$^{-1}$\,kpc$^{-1}$, mutually consistent within their uncertainties. Only degree $3$, which is too rigid to describe the curvature, departs from this (see Appendix~\ref{APP:degree}). The upper limit of $\Rsigmax$\,kpc is set by the sampling of the disc rather than by the fit (Sect.~\ref{sec:param}).

This result should be distinguished from the debate over the behaviour of the rotation curve much further out. \citet{Wang2023} find $V_{\rm c}$ to decline by more than $50$\,km\,s$^{-1}$ between $\approx 15$ and $30$\,kpc, and \citet{Jiao2023} identify a Keplerian decline setting in at $\sim 19$\,kpc, rejecting a flat curve at $3\sigma$ and revising the total mass down to $2.1 \times 10^{11}\,M_\odot$. Those radii lie beyond the reach of our sample. Our measurement concerns the gradient within the disc, over $\Rsigmin < R < \Rsigmax$\,kpc, the interval containing \Fsigrange\% of our sample.

Over the radii common to both, the systematics differ. The two effects that dominate the uncertainty in the outer rotation curve, as reviewed by \citet{MelchiorriRuchika2026}, are the asymmetric drift correction, largest precisely where the measurement is least certain, and the assumptions underlying axisymmetric Jeans modelling, from which cosmological simulations suggest the inferred outer curves may be substantially biased. Neither applies to our determination. Tracers used here are dynamically cold, so no asymmetric drift correction is required, and the curve is obtained by fitting the observed proper motions directly, without recourse to the Jeans equations or to any assumption about the velocity dispersion tensor.

The shape of the rotation curve is informative about the mass distribution of the Galaxy. A declining curve has been discussed in the literature as an indication of a lower total mass than assumed in commonly used models, of a dark matter density profile departing from the standard parameterisations, or of a more centrally concentrated baryonic distribution \citep[e.g.][]{Jiao2023, Wang2023, MelchiorriRuchika2026}. In this work we report the gradient as a measurement. Converting it into a statement about the mass distribution would require assumptions about the baryonic components and the shape of the halo that our data cannot test.

Our tracers occupy a thin layer about the Galactic plane, so the curve constrains the potential in the plane alone. A determination from an older, kinematically lagging population is in progress.

\subsection{The radial solar motion}
\label{sec:degeneracy}

Proper motions constrain only two of the three components of the space velocity. The line-of-sight component is absent from the data but not from the model, where it is fixed implicitly by the assumed velocity field through Eq.~(\ref{eq:vr}), so the fit constrains $U_\odot$ only in combination with whatever systematic radial motion the tracer population possesses. Formal errors on $U_\odot$ from tangential data therefore understate its true uncertainty.

The evidence for this is empirical. \citet{BobylevBajkova2019} solved their kinematic equations for the same $\sim 6000$ OB stars twice, first with line-of-sight velocities alone and then with proper motions alone, and obtained $U_\odot \simeq 9.2$--$9.5$ against $6.1$--$6.3$\,km\,s$^{-1}$, an offset of $\sim 3$\,km\,s$^{-1}$ significant at $4\sigma$. They attribute the discrepancy to the quality of the line-of-sight velocities available for their sample, which may well contribute. The two explanations are not exclusive, and neither can be tested without radial velocities of comparable quality to the astrometry.

Whatever its origin, the instability is apparent in the determinations themselves. Values from OB star samples span several km\,s$^{-1}$, from $5.18 \pm 0.41$\,km\,s$^{-1}$ \citep{PantaleoniGonzalez2024} to $\Uval \pm \Uerr$\,km\,s$^{-1}$ here, in every case by far more than the formal errors of a few tenths. Our own value shifts by $1.8$\,km\,s$^{-1}$ between the two passes of the fit simply on reinstating the stars within $1$\,kpc (Table~\ref{tab:bootstrap}), an effect discussed in Sect.~\ref{sec:local}. The rotation curve is however not appreciably affected. The correlation between $U_\odot$ and $V_{\rm c}(R_\odot)$ across the bootstrap replicates is $\corrUsunVc$, small compared with that of $V_\odot$ ($\corrVsunVc$).

\subsection{Local structure and departures from axisymmetry}
\label{sec:local}

The ALS III sample does not sample the disc uniformly, and the departures from uniformity are mainly correlated with kinematics. 
The nearest few hundred parsecs are dominated by the Gould Belt and the associated system of young OB associations \citep{Zucker2022}, which have a coherent motion of their own. Stars belonging to these structures are a non-negligible fraction of the sample at small heliocentric distances, and their systematic motion biases the inferred solar motion.  \cite{BobylevBajkova2018} find that removing the Gould Belt stars from their own OB sample changes $U_\odot$ and $V_\odot$ by $2.0$ and $2.6$\,km\,s$^{-1}$ while leaving the parameters of the rotation curve essentially unchanged, the same pattern we find on reinstating the nearby stars in the second pass (Table~\ref{tab:bootstrap}). This is distinct from the amplification discussed in Sect.~\ref{sec:pass1}, although it acts in the same direction.

On larger scales, the sample is over-represented towards the nearest spiral arms, a consequence of both the intrinsic clustering of OB stars and the extinction pattern, so that the sky coverage is not uniform in longitude. Beyond their effect on the coverage, the arms leave a kinematic imprint. Our model assumes an axisymmetric velocity field, which the disc is known not to have. Spiral arms drive streaming motions of a few km\,s$^{-1}$, and because our tracers are young and dynamically cold they retain them. For the OB sample most comparable to ours, \citet{BobylevBajkova2019} report radial and tangential perturbation amplitudes of $f_R = 5.1 \pm 1.2$ and $f_\theta = 4.4 \pm 1.4$\,km\,s$^{-1}$. Perturbations of this size are a non-negligible fraction of our residual dispersion and, unlike measurement noise, are spatially correlated. This is one reason the bootstrap uncertainties exceed the formal errors (see Sect.~\ref{sec:residuals}). A related simplification is the assumption of purely circular orbits. \citet{MartinezGarcia2026} find a mean radial gradient significantly different from zero, which we do not fit and which is therefore likewise absorbed into the residuals. We do not attempt to model or measure these non-circular motions here. Characterising the perturbations induced by the spiral arms requires a larger and more precise sample than the present one, and we leave it for Paper~II. They also contribute to $V_{\perp}$, and therefore to the identification of the runaway candidates (Sect.~\ref{sec:robustness}).

A third simplification concerns motion out of the plane. Our model has no vertical component, so the trend of $\delta\mu_{b}$ with $Z$ reported in Sect.~\ref{sec:residuals} is necessarily absorbed into the residuals as well. Its origin is not established by our data. The Galactic warp is a natural candidate at these radii, as is a bending mode of the disc. \citet{Wang2023} also report a marked dependence of $V_{\rm c}$ on $Z$ for $R < 15$\,kpc, which is the range our sample covers. Our sample, confined to a thin layer about the Galactic plane, cannot distinguish among these possibilities.

Since $V_{\rm c}(R)$ is left free in the fit, any smooth radial variation of the mean azimuthal velocity is absorbed into it, so the absence of a trend of the residuals with Galactocentric radius does not by itself show that non-circular motions leave the curve unbiased. The fitted curve is the mean azimuthal velocity of the young population, which equals the circular velocity only to the extent that those motions average out over azimuth. Moreover, a smooth polynomial cannot represent localised departures from axisymmetry on scales of a few hundred parsecs to $\sim 1$\,kpc, such as those driven by the spiral arms, and all three are limitations of the model as implemented.

\subsection{Robustness against kinematic outliers}
\label{sec:robustness}

A fraction of OB stars possess peculiar velocities far in excess of the disc velocity dispersion, having been ejected from their birth sites by dynamical interactions or by the disruption of a binary system \citep{Blaauw1961, Poveda1967}. Runaway stars appear in our analysis as outliers with large values of $V_{\perp}$ (Eq.~\ref{eq:vperp}), and those selected with the fixed threshold number \Noutliers, or \Foutliers\% of the sample.
Since they are retained in the nominal fit, a legitimate concern is whether they distort the fitted rotation curve.

We test this directly by repeating the entire fitting procedure, including the bootstrap, on the sample from which the outliers have been removed. The resulting circular velocity at the solar radius differs from our nominal value by \dVcnorun\,km\,s$^{-1}$, a shift of $\signorun\sigma$ when its uncertainty is estimated from the paired difference of the two fits over the same bootstrap resamples (the two samples share about $90\%$ of their stars). The shift is compensated by the solar motion, with $V_\odot$ changing by $-1.6$\,km\,s$^{-1}$ in the opposite sense, so that $V_{\rm c}(R_\odot) + V_\odot$ is unchanged to within $0.1$\,km\,s$^{-1}$. The rotation curve derived here is therefore not significantly affected by the presence of these objects, which are a minority of the sample.

The outliers themselves constitute the largest sample of spectroscopically confirmed OB runaway candidates compiled to date. Two complementary criteria are applied to select them: a fixed threshold of $V_{\perp} \geq \Vperpthr$\,km\,s$^{-1}$, which yields \Noutliers\ candidates (\Foutliers\% of the sample), and a threshold adapted to the local dispersion of $V_{\perp}$ across the Galactic plane, which accounts for the fact that regions of active star formation and spiral arms are kinematically hotter than the smooth disc, and yields \Nadaptive\ (\Fadaptive\%). These candidates are identified without correcting for large-scale non-circular motions, which are not modelled here (Sect.~\ref{sec:local}). A test in which the large-scale non-circular field is estimated from the data themselves and subtracted from the residuals indicates that the number of candidates would change by between about $3\%$ and $6\%$ for the fixed threshold, depending on how the correction is estimated (we tried several variants, differing in the velocity components corrected and in the stars used to build the correction), and by less than $2\%$ for the adaptive one. It may also affect the membership of individual stars. We therefore regard these counts as candidates, with a systematic uncertainty of that order, and defer their individual identification, kinematic properties and spatial distribution to Paper~II.

\subsection{Comparison with previous determinations}
\label{sec:comparison}

Determinations of $V_{\rm c}(R_\odot)$ are conditioned by the adopted $R_\odot$, which is held fixed in most analyses including ours. The dependence is direct for studies that solve for the angular rotation rate $\Omega_0$, where the linear velocity scales with the adopted distance. We therefore quote the adopted $R_\odot$ alongside each determination below.

\subsubsection{Determinations based on OB stars}

\begin{table*}[t]
\centering
\caption{Determinations of the local rotation parameters from samples of OB stars. The spectral coverage column distinguishes samples selected by effective temperature, which extend to late B, from those restricted to O and early B types. The baseline is the radial extent over which each determination constrains the curve. $V_{\rm c}(R_\odot)$ and the local gradient are as quoted in each work, for the $R_\odot$ given. Entries derived from an angular velocity $\Omega_0$ have been converted using the $R_\odot$ adopted by those authors.}
\label{tab:ob_literature}
\begin{tabular}{lllllrrrr}
\hline\hline
Reference & Gaia & Selection & Spectral & Baseline & $N$ & $R_\odot$ & $V_{\rm c}(R_\odot)$ & ${\rm d}V_{\rm c}/{\rm d}R$\tablefootmark{a} \\
          & data     &           & coverage & (kpc)    &     & (kpc)     & (km\,s$^{-1}$)       & (km\,s$^{-1}$\,kpc$^{-1}$) \\
\hline
\citet{BobylevBajkova2018} & DR2 & spec. & OB     & $\sim 3$ & 495        & 8.0   & $231 \pm 5$            &  -- \\
\citet{BobylevBajkova2019} & DR2 & spec. & OB    & $\sim 4$ & 5335       & 8.0   & $237.6 \pm 4.5$        & -- \\
\citet{GaiaDrimmel2023}    & DR3 & photom. $T_{\rm eff}$ & to B9    &          & 91\,836    & 8.277 & \multicolumn{2}{c}{no curve fitted} \\
\citet{Almannaei2024}      & DR3 & photom. $T_{\rm eff}$ & to B9 & $0.5$ & & 8.178 & $238.30 \pm 2.43$ & $+1.82 \pm 1.17$ \\
\phantom{\citet{Almannaei2024}} & & & & $2$ & & 8.178 & $233.95 \pm 2.24$ & $-3.31 \pm 0.48$ \\
\citet{MartinezGarcia2026} & DR3 & photom. $T_{\rm eff}$ & to B9    & $1$      & 24\,528    & 8.178 & $236.52 \pm 0.11$\tablefootmark{b}      & $-0.73 \pm 0.17$ \\
This work                  & DR3 & spectroscopic & OB & $\Rsigmin$--$\Rsigmax$ & \NfitFinal & \Rsun & $\Vcsun \pm \Vcsunerr$ & $\dVdR \pm \dVdRerr$ \\
\hline
\end{tabular}
\tablefoot{
\tablefoottext{a}{The published values are not all the same quantity. Those of \citet{Almannaei2024} and \citet{MartinezGarcia2026} are mean gradients fitted over the baseline given, whereas ours is the derivative evaluated at $R_\odot$; averaged over $R_\odot \pm 2$\,kpc our curve gives $\slopetwo \pm \slopetwoerr$\,km\,s$^{-1}$\,kpc$^{-1}$. Since the gradient varies with radius, values obtained over different baselines are not directly comparable.}
\tablefoottext{b}{This is the mean azimuthal velocity $V_{\phi,0}$ of their sample at $R_\odot$, which they estimate to differ from the circular velocity by about $1$\,km\,s$^{-1}$ or less through the asymmetric drift. The quoted uncertainty is statistical (MCMC posterior) for fixed solar motion, and does not include the uncertainty of the adopted solar azimuthal velocity ($247.4 \pm 1.4$\,km\,s$^{-1}$). It is therefore not directly comparable with ours.}
}
\end{table*}

Table~\ref{tab:ob_literature} lists the determinations of the rotation curve based on  samples of young massive stars. These are the most direct comparison available, since they
are not affected by the differences in asymmetric drift that complicate comparisons across stellar populations.

\citet{BobylevBajkova2018} analysed 495 OB stars with \textit{Gaia} DR2 astrometry and obtained an angular rotation rate $\Omega_0 = 28.92 \pm 0.39$\,km\,s$^{-1}$\,kpc$^{-1}$, corresponding to $V_0 = 231 \pm 5$\,km\,s$^{-1}$ for their adopted $R_\odot = 8.0$\,kpc. \citet{BobylevBajkova2019} repeated the analysis on the much larger sample of OB stars (up to B2 types) of \citet{Xu2018}, obtaining $\Omega_0 = 29.70 \pm 0.11$\,km\,s$^{-1}$\,kpc$^{-1}$ from a solution using proper motions alone, with Oort constants $A = 16.14 \pm 0.13$ and $B = -13.56 \pm 0.17$\,km\,s$^{-1}$\,kpc$^{-1}$. Because both are expressed as angular quantities, the comparison with our $\Omega_0 = \Omegazero \pm \Omegazeroerr$\,km\,s$^{-1}$\,kpc$^{-1}$ does not depend on the adopted $R_\odot$. We agree with their 2018 determination and differ from the 2019 one by $0.9$\,km\,s$^{-1}$\,kpc$^{-1}$. Our Oort constant $A = \Oorta \pm \Oortaerr$ matches theirs closely, while $B = \Oortb \pm \Oortberr$\,km\,s$^{-1}$\,kpc$^{-1}$ is shallower by about $1$\,km\,s$^{-1}$\,kpc$^{-1}$, reflecting the difference in the local gradient that $A + B$ measures. Their model expands the angular velocity to second order in $(R - R_\odot)$, adequate over the $\sim 4$\,kpc extent of their sample but not a description of the curve across the disc.

The largest OB sample analysed with \textit{Gaia} DR3 to date is that of \citet{GaiaDrimmel2023}, who selected $\sim 580\,000$ stars with $T_{\rm eff} > 10\,000$\,K from the DR3 astrophysical parameters and mapped their velocity field. They do not fit a rotation curve, so the comparison is with their mean azimuthal velocity profile rather than with a fitted gradient, but their sample differs from ours in three respects worth setting out: (i) their selection is based on effective temperature rather than on spectroscopy, which admits subdwarfs and white dwarfs that must be removed by an additional colour-magnitude criterion, along with contaminants from the Magellanic Clouds and from globular clusters; (ii) they adopt the general-purpose photogeometric distances of \citet{BailerJones2021} rather than a prior matched to the OB population; (iii) and the mapping of the velocity field requires line-of-sight velocities, available for only $16\%$ of their sample, so that the velocity field is mapped from a small subset of the astrometric sample. Ours is some forty times smaller than their full sample, but spectroscopically confirmed throughout, and the fit uses the astrometry alone.

Their treatment of the solar height is also instructive. \citet{GaiaDrimmel2023} note that the offset of Sgr\,A$^{*}$ from $b = 0$ implies a residual tilt of $\approx 0\fdg1$ between the Galactic plane and the plane of the Galactic coordinate system, and set $Z_\odot = 0$ on the grounds that the effect is negligible for mapping in the $(X, Y)$ plane. That is a reasonable choice for their purpose, but not for ours, since the tilt terms enter the predicted proper motions directly (see Sect.~\ref{sec:tilt}). Their velocities are likewise not corrected for asymmetric drift, which \citet{Jiao2023} note leaves their OB stars rotating closer to the true circular velocity than corrected tracers of older populations.

\citet{MartinezGarcia2026} fitted a Galactic rotation model, expanded to first order in Galactocentric radius, to the volume-complete sample of O and B stars (including late B types) within $1$\,kpc of \citet{Quintana2025}, obtaining $V_{\phi,0} = 236.52 \pm 0.11$ and ${\rm d}V_\phi/{\rm d}R = -0.73 \pm 0.17$\,km\,s$^{-1}$\,kpc$^{-1}$. Three features of their analysis bear on the comparison: (i) they fix the solar motion to literature values rather than fitting it, adopting $V_{R,\odot} = -10 \pm 1$\,km\,s$^{-1}$ \citep{BlandHawthornGerhard2016} and $V_{\phi,\odot} = 247.4 \pm 1.4$\,km\,s$^{-1}$ \citep{Gravity2019}, which avoids the difficulty discussed in Sect.~\ref{sec:degeneracy} at the cost of propagating any error in the adopted values into their peculiar velocities; (ii) their model uses only the longitudinal component, treating the latitudinal peculiar velocity as a simple offset from its mean, and places the Sun in the Galactic plane. Both simplifications are appropriate within $1$\,kpc, where the tilt terms are negligible, and would not be at the distances spanned by our sample; (iii) the samples themselves barely overlap. Theirs is confined within $1$\,kpc of the Sun, the region we exclude from the first pass precisely because the kinematics there are dominated by local structure (see Sect.~\ref{sec:pass1}), whereas the bulk of ours lies between $1$ and $4$\,kpc.  Our fit nevertheless reproduces, without imposing it, the total azimuthal velocity of the Sun that they adopt: $V_{\rm c}(R_\odot) + V_\odot = 247.2 \pm 1.0$\,km\,s$^{-1}$, where the uncertainty accounts for the anticorrelation between the two terms, against their $247.4 \pm 1.4$\,km\,s$^{-1}$ \citep{Gravity2019}. Their value of $V_{\phi,0}$ then implies $V_\odot \simeq 10.9$\,km\,s$^{-1}$, the same as ours.

The gradients themselves are not mutually consistent within their quoted uncertainties. Part of the spread is expected, since the gradient is not constant with radius and each determination measures it over a different interval, but the baseline alone does not account for it. Averaged over $R_\odot \pm 1$\,kpc our curve gives $-3.48 \pm 0.34$\,km\,s$^{-1}$\,kpc$^{-1}$, much steeper than the $-0.73 \pm 0.17$ of \citet{MartinezGarcia2026} over a comparable baseline, whereas over $R_\odot \pm 2$\,kpc it gives $\slopetwo \pm \slopetwoerr$, within $2\sigma$ of the $-3.31 \pm 0.48$ of \citet{Almannaei2024}. The shallow values obtained within $1$\,kpc are therefore probably related to the differences in the analyses listed above, in particular the confinement of their samples to the region dominated by local structure, rather than to the baseline alone. The positive slope reported by \citet{Almannaei2024} for their $0.5$\,kpc sample is measured over too short a baseline to constrain the gradient, as their own uncertainty indicates. The sign, however, is robust: it is recovered by every OB sample with a baseline of $1$\,kpc or more, and also by tracers not restricted to OB stars, \citet{Akhmetov2024} obtaining $-1.00 \pm 0.07$\,km\,s$^{-1}$\,kpc$^{-1}$ from an all-stars \textit{Gaia} sample. The decline is therefore already detectable in the immediate solar neighbourhood, and our determination extends it over a far longer range.

\subsubsection{Determinations based on other tracers}

Three recent studies measure the circular velocity curve over a range that overlaps ours, using luminous red giants and the Jeans equations: \citet{Eilers2019} from APOGEE spectrophotometric parallaxes, \citet{Ou2024} from an enlarged version of the same sample, and \citet{Zhou2023} from a combined APOGEE and LAMOST sample. Their mean gradients are $-1.7 \pm 0.1$ \citep{Eilers2019} and $-1.83 \pm 0.07$\,km\,s$^{-1}$\,kpc$^{-1}$ \citep{Zhou2023}, to be compared with our $\slopeout \pm \slopeouterr$ over $R_\odot$--$\Rsigmax$\,kpc.

The agreement extends to the curve itself. Table~\ref{tab:rgb_comparison} compares $V_{\rm c}$ at three radii within our significant range. Between $9$ and $10.5$\,kpc our values differ from theirs by less than $2$\,km\,s$^{-1}$, comparable to their own systematic uncertainties of $2$--$4\%$. This is a non-trivial consistency check, since these determinations use a tracer older than ours by two to three orders of magnitude, spectrophotometric rather than astrometric distances, and a Jeans-equation formalism that our method does not employ.

At the solar radius we lie $3$\,km\,s$^{-1}$ above \citet{Ou2024} and \citet{Zhou2023}, an offset present at every polynomial degree we tested. Its sign is what a difference in asymmetric drift would produce. A young, disc-confined tracer measures the circular velocity of a population whose asymmetric drift is negligible, whereas red giants require a correction for it, and \citet{Jiao2023} note that their own correction does not appear sufficient below $10$\,kpc, with OB stars rotating closer to the circular velocity than corrected red giants.

The same consideration applies to comparisons with the maser-based determination of \citet{Reid2014}, the synthesis of \citet{Sofue2020} and the dynamical model of \citet{McMillan2017}, where a residual offset with respect to our value is expected rather than anomalous.

\begin{table}[t]
\centering
\caption{Circular velocity at three Galactocentric radii within our significant range, compared with the red giant determinations of \citet{Ou2024} and \citet{Zhou2023}. Our uncertainties are the bootstrap $1\sigma$ dispersion.}
\label{tab:rgb_comparison}
\begin{tabular}{lrrr}
\hline\hline
$R$ (kpc) & This work & \citet{Ou2024} & \citet{Zhou2023} \\
\hline
 8.2  & $236.2 \pm 1.1$ & 232.8 & 233.2 \\
 9.25 & $232.2 \pm 0.9$ & 230.5 & 232.2 \\
10.25 & $230.0 \pm 0.9$ & 229.4 & 230.3 \\
\hline
\end{tabular}
\end{table}

\section{Conclusions}
\label{sec:conclusions}

We have determined the Galactic rotation curve from \NfitFinal\ spectroscopically confirmed OB stars of the ALS III catalogue, using \textit{Gaia} DR3 astrometry with state-of-the-art systematic corrections and requiring no spectroscopic radial velocities. Our main results are the following.

\begin{enumerate}
  \item Young massive stars yield a rotation curve from five-dimensional astrometry alone. The circular velocity at the solar radius is $V_{\rm c}(R_\odot = \Rsun\,\mathrm{kpc}) = \Vcsun \pm \Vcsunerr$\,km\,s$^{-1}$, a relative precision of $\sim 0.5\%$, with the curve constrained over $\Rsigmin < R < \Rsigmax$\,kpc. 

  \item The curve rises within the solar circle and declines gently beyond $R \approx 8$\,kpc, with a gradient at the solar radius of ${\rm d}V_{\rm c}/{\rm d}R|_{R_\odot} = \dVdR \pm \dVdRerr$\,km\,s$^{-1}$\,kpc$^{-1}$. This gradient has the same sign and a comparable magnitude at every polynomial degree tested. It is measured with a dynamically cold tracer, which requires no asymmetric drift correction, and does not rest on the Jeans-modelling assumptions that dominate the systematics of determinations in the outer disc.

  \item The solar motion of the young disc population is $U_\odot = \Uval \pm \Uerr$, $V_\odot = \Vval \pm \Verr$ and  $W_\odot = \Wval \pm \Werr$\,km\,s$^{-1}$, with Oort constants  $A = \Oorta \pm \Oortaerr$ and $B = \Oortb \pm \Oortberr$\,km\,s$^{-1}$\,kpc$^{-1}$.

  \item Fits based on proper motions alone constrain the solar radial motion only in combination with the systematic line-of-sight motion of the tracer population, so that $U_\odot$ carries an uncertainty that its formal error does not capture. This does not propagate appreciably into the rotation curve itself.

  \item A two-pass fitting scheme, in which stars closer than $1$\,kpc are excluded from the first pass and reinstated in the second, separates the constraint on the large-scale rotation curve from the local velocity field. Reinstating the nearby stars shifts $U_\odot$ and $V_\odot$ by $+1.8$ and $+2.4$\,km\,s$^{-1}$ respectively but $W_\odot$ by only $-0.4$\,km\,s$^{-1}$, as expected if the effect originates in structures whose coherent motion lies in the disc plane.

  \item The residuals show no trend with Galactocentric radius, though a trend with height above the plane is present in the latitudinal component. A minority of stars depart from the curve by far more than the disc velocity dispersion. Selected by their peculiar tangential velocity $V_{\perp}$, they form the largest sample of OB runaway candidates compiled to date: \Noutliers\ (\Foutliers\% of the sample) with a fixed threshold, and \Nadaptive\ (\Fadaptive\%) with a threshold adapted to the local dispersion of $V_{\perp}$. These counts carry a systematic uncertainty of up to about $6\%$, owing to the non-circular motions that the model does not describe. Removing the stars selected with the fixed threshold and refitting changes $V_{\rm c}(R_\odot)$ by only \dVcnorun\,km\,s$^{-1}$ ($\signorun\sigma$), with an opposite change in $V_\odot$ that leaves $V_{\rm c}(R_\odot) + V_\odot$ unchanged, so the curve reported here is not significantly affected by their presence.

\end{enumerate}

The principal limitation of this analysis is the absence of spectroscopic radial velocities, which is what renders the solar radial motion poorly constrained. Radial velocities for a representative subsample would resolve this directly, and the prospects for obtaining them are good: both WEAVE \citep{Jin2024} and 4MOST \citep{deJong2019} include surveys targeting Galactic massive stars, and will deliver radial velocities for large numbers of OB stars. \textit{Gaia} DR4 will reduce the astrometric uncertainties further and extend the sample, and an independent determination from an older, kinematically lagging tracer population is in progress, which will allow the outer slope reported here to be cross-validated. The characterisation of the spiral-arm perturbations and other non-circular motions that our axisymmetric model does not describe, and the individual identification and analysis of the candidate runaway OB stars, are deferred to Paper~II, which will be based on \textit{Gaia} DR4.

\begin{acknowledgements}

This work has made use of data from the European Space Agency (ESA) mission \textit{Gaia} (\url{https://www.cosmos.esa.int/gaia}), processed by the Gaia DPAC. This research has made use of the SIMBAD database and the VizieR catalogue access tool (CDS, Strasbourg, France; \citealt{Wenger2000, Ochsenbein2000}). This work is part of grant CEX2025-001609-S, awarded to the Instituto de Astrofísica de Canarias under the Severo Ochoa Centre of Excellence program and funded by MICIU/AEI/10.13039/501100011033. SRB acknowledges the support of the Viera y Clavijo program funded by the Agencia Canaria de Investigación, Innovación y Sociedad de la Información del Gobierno de Canarias and the Universidad de la Laguna, and from the State Research Agency (AEI) of the Spanish Ministry of Science and Innovation (MICIN) and the European Regional Development Fund, FEDER under grant PID2024-159329NB-C21. MPG was co-Funded by the European Union (ERC, ISM-FLOW, 101055318) and acknowledges funding from the European Research Council (ERC) under the European Union’s Horizon 2020 research and innovation programme (Grant agreement No. 851435).

\end{acknowledgements}

\bibliographystyle{aa}
\bibliography{references_chev3}

@ARTICLE{PantaleoniGonzalez2025,
  author  = {{Pantaleoni Gonz\'alez}, M. and {Ma\'iz Apell\'aniz}, J. and
             {Barb\'a}, R.~H. and {Cameron Reed}, B. and {Berlanas}, S.~R. and
             {Parras Rico}, A. and {Bodaghee}, A.},
  title   = "{The Alma catalogue of OB stars -- III. A cross-match with Gaia DR3
              and an extension based on new spectral classifications}",
  journal = {\mnras},
  year    = 2025,
  volume  = 543,
  number  = 1,
  pages   = {63--82},
  doi     = {10.1093/mnras/staf1409}
}

@Article{Maizetal08a,
 author        = {Ma{\'\i}z Apell{\'a}niz, J. and Alfaro, E.~J. and Sota, A.},
 title         = {Accurate distances to nearby massive stars with the new reduction of the Hipparcos raw data},
 journal       = {arXiv:0804.2553},
 year          = {2008},
 month         = {apr},
 adsurl        = {https://adsabs.net/abs/2008arXiv0804.2553M}}

@ARTICLE{MaizApellaniz2022,
       author = {{Ma{\'\i}z Apell{\'a}niz}, J.},
        title = "{An estimation of the Gaia EDR3 parallax bias from stellar clusters and Magellanic Clouds data}",
      journal = {\aap},
         year = 2022,
        month = jan,
       volume = {657},
          eid = {A130},
        pages = {A130},
          doi = {10.1051/0004-6361/202142365},
archivePrefix = {arXiv},
       eprint = {2110.01475},
 primaryClass = {astro-ph.IM},
       adsurl = {https://ui.adsabs.harvard.edu/abs/2022A&A...657A.130M}
}

@ARTICLE{MaizApellaniz2021,
       author = {{Ma{\'\i}z Apell{\'a}niz}, J. and {Pantaleoni Gonz{\'a}lez}, M. and {Barb{\'a}}, R.~H.},
        title = "{Validation of the accuracy and precision of Gaia EDR3 parallaxes with globular clusters}",
      journal = {\aap},
         year = 2021,
        month = may,
       volume = {649},
          eid = {A13},
        pages = {A13},
          doi = {10.1051/0004-6361/202140418},
archivePrefix = {arXiv},
       eprint = {2101.10206},
 primaryClass = {astro-ph.IM},
       adsurl = {https://ui.adsabs.harvard.edu/abs/2021A&A...649A..13M}
}

@article{MaizApellaniz2001,
doi = {10.1086/320399},
url = {https://doi.org/10.1086/320399},
year = {2001},
month = {may},
publisher = {},
volume = {121},
number = {5},
pages = {2737},
author = {Ma\'iz Apell\'aniz, Jesús},
title = {The Spatial Distribution of O-B5 Stars in the Solar Neighborhood as Measured by Hipparcos*},
journal = {The Astronomical Journal}
}

@ARTICLE{CantatGaudin2021,
  author  = {{Cantat-Gaudin}, T. and {Brandt}, T.~D.},
  title   = "{Characterising and correcting the proper motion bias of the bright
              Gaia EDR3 sources}",
  journal = {\aap},
  year    = 2021,
  volume  = 649,
  pages   = {A124},
  doi     = {10.1051/0004-6361/202140807}
}

@ARTICLE{Lindegren2021,
  author  = {{Lindegren}, L. and {Bastian}, U. and {Biermann}, M. and others},
  title   = "{Gaia Early Data Release 3: Parallax bias versus magnitude, colour,
              and position}",
  journal = {\aap},
  year    = 2021,
  volume  = 649,
  pages   = {A4},
  doi     = {10.1051/0004-6361/202039653}
}

@MISC{ButkevichLindegren2022,
  author       = {{Butkevich}, A.~G. and {Lindegren}, L.},
  title        = "{Gaia DR3 Documentation, Section 4.1.7: Transformation of
                   astrometric data and error propagation}",
  howpublished = {Gaia DR3 Documentation, European Space Agency and
                   Gaia Data Processing and Analysis Consortium},
  year         = 2022
}

@ARTICLE{BailerJones2021,
  author  = {{Bailer-Jones}, C.~A.~L. and {Rybizki}, J. and {Fouesneau}, M. and
             {Demleitner}, M. and {Andrae}, R.},
  title   = "{Estimating Distances from Parallaxes. V. Geometric and
              Photogeometric Distances to 1.47 Billion Stars in Gaia EDR3}",
  journal = {\aj},
  year    = 2021,
  volume  = 161,
  pages   = {147},
  doi     = {10.3847/1538-3881/abd806}
}

@ARTICLE{LutzKelker1973,
  author  = {{Lutz}, T.~E. and {Kelker}, D.~H.},
  title   = "{On the Use of Trigonometric Parallaxes for the Calibration of
              Luminosity Systems: Theory}",
  journal = {\pasp},
  year    = 1973,
  volume  = 85,
  pages   = {573},
  doi     = {10.1086/129506}
}

@ARTICLE{Gravity2019,
  author  = {{GRAVITY Collaboration} and {Abuter}, R. and {Amorim}, A. and others},
  title   = "{A geometric distance measurement to the Galactic center black hole
              with 0.3\% uncertainty}",
  journal = {\aap},
  year    = 2019,
  volume  = 625,
  pages   = {L10},
  doi     = {10.1051/0004-6361/201935656}
}

@ARTICLE{Anderson2019,
       author = {{Anderson}, L.~D. and {Wenger}, Trey V. and {Armentrout}, W.~P. and {Balser}, Dana S. and {Bania}, T.~M.},
        title = "{A Galactic Plane Defined by the Milky Way H II Region Distribution}",
      journal = {\apj},
         year = 2019,
        month = feb,
       volume = {871},
       number = {2},
          eid = {145},
        pages = {145},
          doi = {10.3847/1538-4357/aaf571},
archivePrefix = {arXiv},
       eprint = {1812.02244},
 primaryClass = {astro-ph.GA},
       adsurl = {https://ui.adsabs.harvard.edu/abs/2019ApJ...871..145A}
}

@MISC{lmfit,
  author       = {{Newville}, M. and {Stensitzki}, T. and {Allen}, D.~B. and
                  {Ingargiola}, A.},
  title        = {LMFIT: Non-Linear Least-Square Minimization and Curve-Fitting
                  for Python},
  year         = 2014,
  doi          = {10.5281/zenodo.11813},
  howpublished = {Zenodo}
}

@ARTICLE{BobylevBajkova2018,
       author = {{Bobylev}, V.~V. and {Bajkova}, A.~T.},
        title = "{Kinematics of the Galaxy from OB Stars with Data from the Gaia DR2 Catalogue}",
      journal = {Astronomy Letters},
         year = 2018,
        month = nov,
       volume = {44},
       number = {11},
        pages = {676-687},
          doi = {10.1134/S1063773718110026},
archivePrefix = {arXiv},
       eprint = {1809.10512},
 primaryClass = {astro-ph.GA},
       adsurl = {https://ui.adsabs.harvard.edu/abs/2018AstL...44..676B}
}

@ARTICLE{BobylevBajkova2019,
       author = {{Bobylev}, V.~V. and {Bajkova}, A.~T.},
        title = "{Galactic Rotation Based on OB Stars from the Gaia DR2 Catalogue}",
      journal = {Astronomy Letters},
         year = 2019,
        month = jun,
       volume = {45},
       number = {6},
        pages = {331-340},
          doi = {10.1134/S106377371906001X},
archivePrefix = {arXiv},
       eprint = {1906.10151},
 primaryClass = {astro-ph.GA},
       adsurl = {https://ui.adsabs.harvard.edu/abs/2019AstL...45..331B}
}

@ARTICLE{Xu2018,
  author  = {{Xu}, Y. and {Bian}, S.~B. and {Reid}, M.~J. and others},
  title   = "{Delineating the Milky Way's spiral structure with OB-type stars}",
  journal = {\aap},
  year    = 2018,
  volume  = 616,
  pages   = {L15},
  doi     = {10.1051/0004-6361/201833407}
}

@ARTICLE{Almannaei2024,
  author  = {{Almannaei}, A.~S. and {Kawata}, D. and {Baba}, J. and
             {Hunt}, J.~A.~S. and {Seabroke}, G. and {Yan}, Z.},
  journal = {\mnras},
  year    = 2024,
  volume  = 529,
  pages   = {1035},
  doi     = {10.1093/mnras/stae158}
}

@ARTICLE{Akhmetov2024,
  author  = {{Akhmetov}, V.~S. and {Bucciarelli}, B. and {Crosta}, M. and
             {Lattanzi}, M.~G. and {Spagna}, A. and {Re Fiorentin}, P. and
             {Bannikova}, E.~Y.},
  journal = {\mnras},
  year    = 2024,
  volume  = 530,
  pages   = {710},
  doi     = {10.1093/mnras/stae772}
}

@ARTICLE{MartinezGarcia2026,
  author  = {{Mart\'inez Garc\'ia}, J. and {Wright}, N.~J. and {Quintana}, A.~L.},
  title   = "{Runaway OB stars within 1 kpc of the Sun}",
  journal = {\mnras},
  year    = 2026,
  volume  = 550,
  number  = 2,
  pages   = {stag1239},
  doi     = {10.1093/mnras/stag1239}
}

@ARTICLE{Quintana2025,
  author  = {{Quintana}, A.~L. and {Wright}, N.~J. and {Mart\'inez Garc\'ia}, J.},
  title   = "{A census of OB stars within 1 kpc and the star formation and core
              collapse supernova rates of the Milky Way}",
  journal = {\mnras},
  year    = 2025,
  volume  = 538,
  number  = 3,
  pages   = {1367--1383},
  doi     = {10.1093/mnras/staf083}
}

@MASTERSTHESIS{PantaleoniGonzalez2024,
author = {Pantaleoni González, Michelangelo and Maíz Apellániz, Jesús and Gil de Paz, Armando},
year = {2024},
month = {06},
pages = {},
school= {UCM},
title = {Galactic structure from the kinematical analysis of massive stars with Gaia - Thesis for the Master in Astrophysics},
doi = {10.13140/RG.2.2.31184.32004}
}

@ARTICLE{Reid2014,
  author  = {{Reid}, M.~J. and {Menten}, K.~M. and {Brunthaler}, A. and others},
  title   = "{Trigonometric Parallaxes of High Mass Star Forming Regions: The
              Structure and Kinematics of the Milky Way}",
  journal = {\apj},
  year    = 2014,
  volume  = 783,
  pages   = {130},
  doi     = {10.1088/0004-637X/783/2/130}
}

@ARTICLE{Sofue2020,
  author  = {{Sofue}, Y.},
  title   = "{Rotation Curve of the Milky Way and the Dark Matter Density}",
  journal = {Galaxies},
  year    = 2020,
  volume  = 8,
  number  = 2,
  pages   = {37},
  doi     = {10.3390/galaxies8020037}
}

@ARTICLE{Sofue2009,
  author  = {{Sofue}, Y. and {Honma}, M. and {Omodaka}, T.},
  title   = "{Unified Rotation Curve of the Galaxy}",
  journal = {\pasj},
  year    = 2009,
  volume  = 61,
  pages   = {227},
  doi     = {10.1093/pasj/61.2.227}
}

@ARTICLE{McMillan2017,
  author  = {{McMillan}, P.~J.},
  title   = "{The mass distribution and gravitational potential of the Milky Way}",
  journal = {\mnras},
  year    = 2017,
  volume  = 465,
  pages   = {76},
  doi     = {10.1093/mnras/stw2759}
}

@ARTICLE{Jiao2023,
  author  = {{Jiao}, Y. and {Hammer}, F. and {Wang}, H. and {Wang}, J. and
             {Amram}, P. and {Chemin}, L. and {Yang}, Y.},
  title   = "{Detection of the Keplerian decline in the Milky Way rotation curve}",
  journal = {\aap},
  year    = 2023,
  volume  = 678,
  pages   = {A208},
  doi     = {10.1051/0004-6361/202347513}
}

@ARTICLE{Wang2023,
  author  = {{Wang}, H.-F. and {Chrob\'akov\'a}, {\v Z}. and
             {L\'opez-Corredoira}, M. and {Sylos Labini}, F.},
  title   = "{Mapping the Milky Way Disk with Gaia DR3: 3D Extended Kinematic
              Maps and Rotation Curve to $\approx$30 kpc}",
  journal = {\apj},
  year    = 2023,
  volume  = 942,
  number  = 1,
  pages   = {12},
  doi     = {10.3847/1538-4357/aca27c}
}

@ARTICLE{BlandHawthornGerhard2016,
  author  = {{Bland-Hawthorn}, J. and {Gerhard}, O.},
  title   = "{The Galaxy in Context: Structural, Kinematic, and Integrated
              Properties}",
  journal = {\araa},
  year    = 2016,
  volume  = 54,
  pages   = {529},
  doi     = {10.1146/annurev-astro-081915-023441}
}

@ARTICLE{Zucker2022,
  author  = {{Zucker}, C. and {Goodman}, A.~A. and {Alves}, J. and others},
  title   = "{Star formation near the Sun is driven by expansion of the
              Local Bubble}",
  journal = {Nature},
  year    = 2022,
  volume  = 601,
  pages   = {334},
  doi     = {10.1038/s41586-021-04286-5}
}

@ARTICLE{Jin2024,
  author  = {{Jin}, S. and {Trager}, S.~C. and {Dalton}, G.~B. and others},
  title   = "{The wide-field, multiplexed, spectroscopic facility WEAVE:
              Survey design, overview, and simulated implementation}",
  journal = {\mnras},
  year    = 2024,
  volume  = 530,
  pages   = {2688},
  doi     = {10.1093/mnras/stad557}
}

@ARTICLE{deJong2019,
  author  = {{de Jong}, R.~S. and {Agertz}, O. and {Berbel}, A.~A. and others},
  title   = "{4MOST: Project overview and information for the First Call for
              Proposals}",
  journal = {The Messenger},
  year    = 2019,
  volume  = 175,
  pages   = {3},
  doi     = {10.18727/0722-6691/5117}
}

@ARTICLE{Wenger2000,
  author  = {{Wenger}, M. and {Ochsenbein}, F. and {Egret}, D. and others},
  title   = "{The SIMBAD astronomical database. The CDS reference database for
              astronomical objects}",
  journal = {\aaps},
  year    = 2000,
  volume  = 143,
  pages   = {9},
  doi     = {10.1051/aas:2000332}
}

@ARTICLE{Ochsenbein2000,
  author  = {{Ochsenbein}, F. and {Bauer}, P. and {Marcout}, J.},
  title   = "{The VizieR database of astronomical catalogues}",
  journal = {\aaps},
  year    = 2000,
  volume  = 143,
  pages   = {23},
  doi     = {10.1051/aas:2000169}
}

@ARTICLE{PaperII,
  author  = {{Berlanas}, S.~R. and {Pantaleoni Gonz\'alez}, M. and
             {Ma\'iz Apell\'aniz}, J.},
  title   = "{Runaway OB stars in the Galactic disc: identification, kinematics
              and validation using the ALS III catalogue}",
  journal = {\aap},
  year    = 2026,
  note    = {submitted (Paper II)}
}

@ARTICLE{Morgan1951,
  author  = {{Morgan}, W.~W.},
  title   = "{Application of the Principle of Natural Groups to the
              Classification of Stellar Spectra}",
  journal = {Publications of the Michigan Observatory},
  year    = 1951,
  volume  = 10,
  pages   = {33}
}

@Inproceedings{Maizetal26a,
 author        = {Ma{\'\i}z Apell{\'a}niz, Jes{\'u}s and Negueruela, Ignacio and Caballero, Jos{\'e} A.},
 title         = {Spectral classification},
 booktitle     = {Encyclopedia of Astrophysics},
 year          = 2026,
 volume        = {2},
 month         = jan,
 pages         = {43-84},
 doi           = {10.1016/B978-0-443-21439-4.00057-2},
 adsurl        = {https://ui.adsabs.harvard.edu/abs/2026enap....2...43M}}

@ARTICLE{Schwarz1978,
  author  = {{Schwarz}, G.},
  title   = "{Estimating the Dimension of a Model}",
  journal = {Annals of Statistics},
  year    = 1978,
  volume  = 6,
  pages   = {461--464},
  doi     = {10.1214/aos/1176344136}
}

@ARTICLE{Blaauw1961,
  author  = {{Blaauw}, A.},
  title   = "{On the origin of the O- and B-type stars with high velocities
              (the run-away stars), and some related problems}",
  journal = {\bain},
  year    = 1961,
  volume  = 15,
  pages   = {265}
}

@ARTICLE{Poveda1967,
  author  = {{Poveda}, A. and {Ruiz}, J. and {Allen}, C.},
  title   = "{Run-away Stars as the Result of the Gravitational Collapse of
              Proto-stellar Clusters}",
  journal = {Boletin de los Observatorios Tonantzintla y Tacubaya},
  year    = 1967,
  volume  = 4,
  pages   = {86}
}

@ARTICLE{Eilers2019,
  author  = {{Eilers}, A.-C. and {Hogg}, D.~W. and {Rix}, H.-W. and
             {Ness}, M.~K.},
  title   = "{The Circular Velocity Curve of the Milky Way from 5 to 25 kpc}",
  journal = {\apj},
  year    = 2019,
  volume  = 871,
  pages   = {120},
  doi     = {10.3847/1538-4357/aaf648}
}

@ARTICLE{MelchiorriRuchika2026,
  author  = {{Melchiorri}, A. and {Ruchika}},
  title   = "{The Rotation Curve of the Milky Way: State of the Art, the
              Keplerian Decline Debate, and Implications for Dark Matter}",
  journal = {New Astronomy Reviews},
  year    = 2026,
  volume  = 103,
  eprint  = {2608.10189}
}

@ARTICLE{Ou2024,
  author  = {{Ou}, X. and {Eilers}, A.-C. and {Necib}, L. and {Frebel}, A.},
  title   = "{The dark matter profile of the Milky Way inferred from its
              circular velocity curve}",
  journal = {\mnras},
  year    = 2024,
  volume  = 528,
  pages   = {693},
  doi     = {10.1093/mnras/stae034}
}

@ARTICLE{GaiaDrimmel2023,
  author  = {{Gaia Collaboration} and {Drimmel}, R. and {Romero-G\'omez}, M. and others},
  title   = "{Gaia Data Release 3: Mapping the asymmetric disc of the Milky Way}",
  journal = {\aap},
  year    = 2023,
  volume  = 674,
  pages   = {A37},
  doi     = {10.1051/0004-6361/202243797}
}

@ARTICLE{Zhou2023,
  author  = {{Zhou}, Y. and {Li}, X. and {Huang}, Y. and {Zhang}, H.},
  title   = "{The Circular Velocity Curve of the Milky Way from 5 to 25 kpc
              Using Luminous Red Giant Branch Stars}",
  journal = {\apj},
  year    = 2023,
  volume  = 946,
  pages   = {73},
  doi     = {10.3847/1538-4357/acb545}
}

  
\begin{appendix}

\section{Coordinate transformations and full model equations}
\label{APP:transf}

We collect here the complete expressions used in the fit. The derivation follows \citet{PantaleoniGonzalez2024}, to which we refer for the intermediate steps. We reproduce the result in full because the terms arising from the solar height are omitted in most published treatments. 

\subsection{Reference frames}

Let $\{\hat{x}, \hat{y}, \hat{z}\}$ be the heliocentric Cartesian basis of the Galactic coordinate system, with $\hat{x}$ pointing towards the Galactic centre, $\hat{y}$ in the direction of Galactic rotation and $\hat{z} = \hat{x} \times \hat{y}$ towards the North Galactic Pole. The spherical basis attached to a star at $(l, b)$ is
\begin{equation}
\left\{
\begin{aligned}
\hat{r} &= \cos b \cos l\, \hat{x} + \cos b \sin l\, \hat{y} + \sin b\, \hat{z} \\
\hat{b} &= -\sin b \cos l\, \hat{x} - \sin b \sin l\, \hat{y} + \cos b\, \hat{z} \\
\hat{l} &= -\sin l\, \hat{x} + \cos l\, \hat{y}
\end{aligned}
\right.
\label{eq:spherical_basis}
\end{equation}
and the velocity of a star relative to the Sun decomposes as
\begin{equation}
\vec{v}^{\,\star}_{\odot} = V_{r}\,\hat{r} + r\mu_{b}\,\hat{b} + r\mu_{l}^{*}\,\hat{l} ,
\label{eq:vstar}
\end{equation}
with $\mu_{l}^{*} \equiv \mu_{l}\cos b$.

\subsection{The tilted Galactic plane}

Because the Sun lies at a height $Z_\odot$ above the Galactic midplane while the origin of the Galactic coordinate system points at the Galactic centre, the $b = 0$ plane is not parallel to the midplane. Writing $s = Z_\odot/R_\odot$, the normal to the true Galactic plane expressed in the star's spherical basis is
\begin{align}
\hat{z}_{\rm G} = & \left(\sqrt{1-s^{2}}\sin b - s\cos b\cos l\right)\hat{r}  \nonumber \\
              &  + \left(\sqrt{1-s^{2}}\cos b + s\sin b\cos l\right)\hat{b}
                + s\sin l\, \hat{l} 
\label{eq:znormal}
\end{align}
Projecting the heliocentric position vector onto the Galactic plane with $\vec{r}^{\;*} = \vec{r} - (\vec{r}\cdot\hat{z}_{\rm G})\hat{z}_{\rm G}$ and applying the cosine rule to the triangle formed by $R$, $R_{\rm c}$ and $r^{*}$ yields Eq.~(\ref{eq:Rgal}) of the main text, which we repeat here in the form in which it is implemented,
\begin{align}
R^{2} = \; & R_{\rm c}^{2} + r^{2}c_{b}^{2} - 2\sqrt{1-s^{2}}\,R_{\rm c}\,r\,c_{b}c_{l}
             - 2 s R_{\rm c}\, r\, s_{b} \nonumber \\
           & + s^{2} r^{2}\left(s_{b}^{2} - c_{b}^{2} c_{l}^{2}\right)
             + 2 s \sqrt{1-s^{2}}\, r^{2} c_{b} s_{b} c_{l} ,
\label{eq:Rgal_app}
\end{align}
where we abbreviate $s_{l} = \sin l$, $c_{l} = \cos l$, $s_{b} = \sin b$, $c_{b} = \cos b$. Equivalently, this is $R^{2} = X^{2} + Y^{2}$ with the Galactocentric Cartesian coordinates of the star given by
\begin{equation}
\left\{
\begin{aligned}
X &= R_{\rm c} - r\sqrt{1-s^{2}}\,c_{b}c_{l} - r s\, s_{b} \\
Y &= -r\, c_{b} s_{l} \\
Z &= Z_\odot - r s\, c_{b}c_{l} + r\sqrt{1-s^{2}}\, s_{b} .
\end{aligned}
\right.
\label{eq:cartesian}
\end{equation}

\subsection{Velocity field and predicted observables}

Assuming that the mean velocity field of the disc is cylindrical, so that the velocities are parallel to the Galactic plane and perpendicular to the cylindrical radius vector, the velocity of a star on a circular orbit is $\vec{V}(R) = V_{\rm c}(R)\,\hat{\varphi}$ with $\hat{\varphi} = -\hat{z}_{\rm G} \times \hat{R}$. Composing the motion of the star about the Galactic centre with the motion of the Sun, and projecting the difference onto the basis of Eq.~(\ref{eq:spherical_basis}), gives for stars with no peculiar velocity
\begin{align}
\mu_{l}^{\rm th} = \frac{\kappa}{r}\Big[\, & U_\odot s_{l} + \left(\frac{R_{\rm c}}{R}V_{\rm c}(R) - V_\odot - V_{\rm c,\odot}\right)c_{l} \nonumber \\
  & - \frac{r}{R}V_{\rm c}(R)\left(\sqrt{1-s^{2}}\,c_{b} + s\, s_{b} c_{l}\right)\Big] , \label{eq:pml_app}\\[4pt]
\mu_{b}^{\rm th} = \frac{\kappa}{r}\Big[\, & U_\odot s_{b} c_{l} - \left(\frac{R_{\rm c}}{R}V_{\rm c}(R) - V_\odot - V_{\rm c,\odot}\right)s_{b} s_{l} \nonumber \\
  & - W_\odot c_{b} + \frac{r\, Z_\odot}{R\, R_\odot}V_{\rm c}(R)\, s_{l} \Big] , \label{eq:pmb_app}\\[4pt]
V_{r}^{\rm th} = \; & -U_\odot c_{b} c_{l} + \left(\frac{R_{\rm c}}{R}V_{\rm c}(R) - V_\odot - V_{\rm c,\odot}\right)c_{b} s_{l} - W_\odot s_{b} . \label{eq:vr_app}
\end{align}
The last term of Eq.~(\ref{eq:pmb_app}) is the tilt contribution to the latitudinal proper motion, and it is absent from treatments that place the Sun in the Galactic plane. Written in terms of the angular velocity $\omega \equiv V_{\rm c}(R)/R$ it takes the compact form $\omega\, s\, r\, s_{l}$ used in Eq.~(\ref{eq:pmb}).

\subsection{Numerical constants}

The conversion factor from km\,s$^{-1}$\,pc$^{-1}$ to mas\,yr$^{-1}$ is $\kappa = 0.210949$, the reciprocal of the familiar $4.740471$\,km\,s$^{-1}$ per mas\,yr$^{-1}$\,kpc.

\subsection{Neglected corrections}

Two further refinements are available but are not applied here. First, the origin of the IAU Galactic coordinate system points at the radio source Sgr\,A rather than at Sgr\,A$^{*}$, which we take to lie at the barycentre of the Galaxy. The true Galactic centre is therefore offset by $\approx 4\farcm3$ from the coordinate origin, introducing an additional tilt and a roll of the coordinate frame \citep{Anderson2019}. These are roughly half the size of the tilt due to the solar height. Second, the assumption of a flat Galactic plane should itself be revisited at this level of detail, since the disc is warped. The vertical trend we find in the residuals (Sect.~\ref{sec:residuals}) may be related, and is discussed in Sect.~\ref{sec:local}. We regard both refinements as beyond the accuracy warranted by the present data, and note them so that the scope of the model is explicit.

\section{Choice of the polynomial degree}
\label{APP:degree}

Polynomials of degree $n = 3$ to $6$ were fitted to the same sample following the identical procedure, with \Nboot\ bootstrap replicates in each case. Table~\ref{tab:degree} collects the results.

\begin{table*}[t]
\centering
\caption{Dependence of the results on the polynomial degree. $\Delta$BIC is referred to the best-fitting model, $n = 5$. The fifth column gives the gradient averaged over $R_\odot$--$\Rsigmax$\,kpc, the quantity comparable with determinations fitted over a long baseline, as opposed to the derivative at $R_\odot$ in the fourth. Uncertainties are the $1\sigma$ bootstrap dispersion. The last column gives the largest relative uncertainty among the individual coefficients of each fit. }
\label{tab:degree}
\begin{tabular}{crrrrr}
\hline\hline
$n$ & $\Delta$BIC & $V_{\rm c}(R_\odot)$ & ${\rm d}V_{\rm c}/{\rm d}R|_{R_\odot}$ & $\langle {\rm d}V_{\rm c}/{\rm d}R \rangle$ & max.\ rel.\ err. \\
    &             & (km\,s$^{-1}$)       & (km\,s$^{-1}$\,kpc$^{-1}$)             & (km\,s$^{-1}$\,kpc$^{-1}$)                  & (\%) \\
\hline
3 & $+7.7$  & $236.98 \pm 1.09$ & $-3.19 \pm 0.27$ & $-3.75 \pm 0.38$ & 19 \\
4 & $+3.4$  & $236.25 \pm 1.09$ & $-3.81 \pm 0.38$ & $-2.42 \pm 0.59$ & 25 \\
5 & $0$     & $236.17 \pm 1.26$ & $-3.73 \pm 0.44$ & $-2.44 \pm 0.57$ & 254 \\
6 & $+9.9$  & $236.00 \pm 1.20$ & $-4.07 \pm 0.60$ & $-2.06 \pm 0.69$ & 353 \\
\hline
\end{tabular}
\end{table*}

Two results are robust across the four fits. $V_{\rm c}(R_\odot)$ varies by only $1$\,km\,s$^{-1}$ from end to end, within the bootstrap uncertainty of any individual fit, so the circular velocity we report does not depend on this choice. The local gradient is negative at every degree, between $-3.2$ and $-4.1$\,km\,s$^{-1}$\,kpc$^{-1}$ and in each case significant at more than $6\sigma$. The decline reported in Sect.~\ref{sec:outer} is a property of the data and not of the flexibility granted to the model.

The choice among degrees is less clear-cut than these numbers suggest, and we set it out explicitly. Degree $6$ is excluded on both counts: it is disfavoured by the BIC and two of its coefficients are consistent with zero. Degree $3$ is disfavoured by the BIC ($\Delta{\rm BIC} = 7.7$) and gives a somewhat shallower local gradient than the other three, although the difference is within the uncertainties. Between degrees $4$ and $5$ the evidence is genuinely mixed. The BIC favours degree $5$ by $3.4$, but it is only indicative, since the residuals are not independent (Sect.~\ref{sec:param}), and degree $5$ achieves it with a coefficient determined to no better than $250\%$, while the two give values of $V_{\rm c}(R_\odot)$ and of the gradient that agree well within their uncertainties. We adopt $n = \polydeg$ as the highest degree at which every coefficient is individually significant, while noting that degree $5$ would be the formally preferred model and that nothing in our conclusions would change had we adopted it.

The mean gradient over $R_\odot$--$\Rsigmax$\,kpc behaves in the same way. It is $-2.42$, $-2.44$ and $-2.06$\,km\,s$^{-1}$\,kpc$^{-1}$ at degrees $4$, $5$ and $6$, mutually consistent within uncertainties of about $0.6$, while degree $3$ gives $-3.75 \pm 0.38$ and departs from the others. The total decrease of the curve across the same interval, obtained by integrating the fitted gradient, is less stable between models that describe the data equally well, and we do not quote it.

\begin{figure*}[t]
\centering
\includegraphics[width=\textwidth]{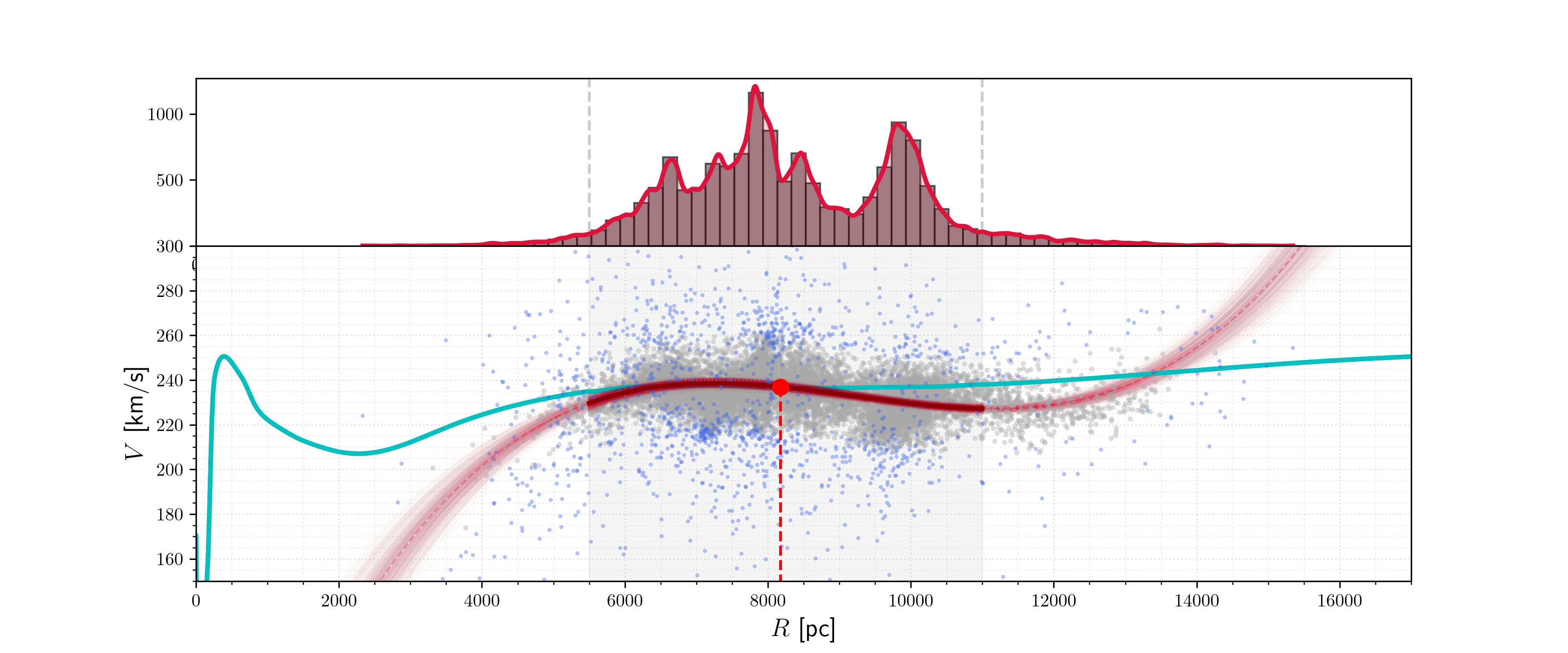}
\includegraphics[width=\textwidth]{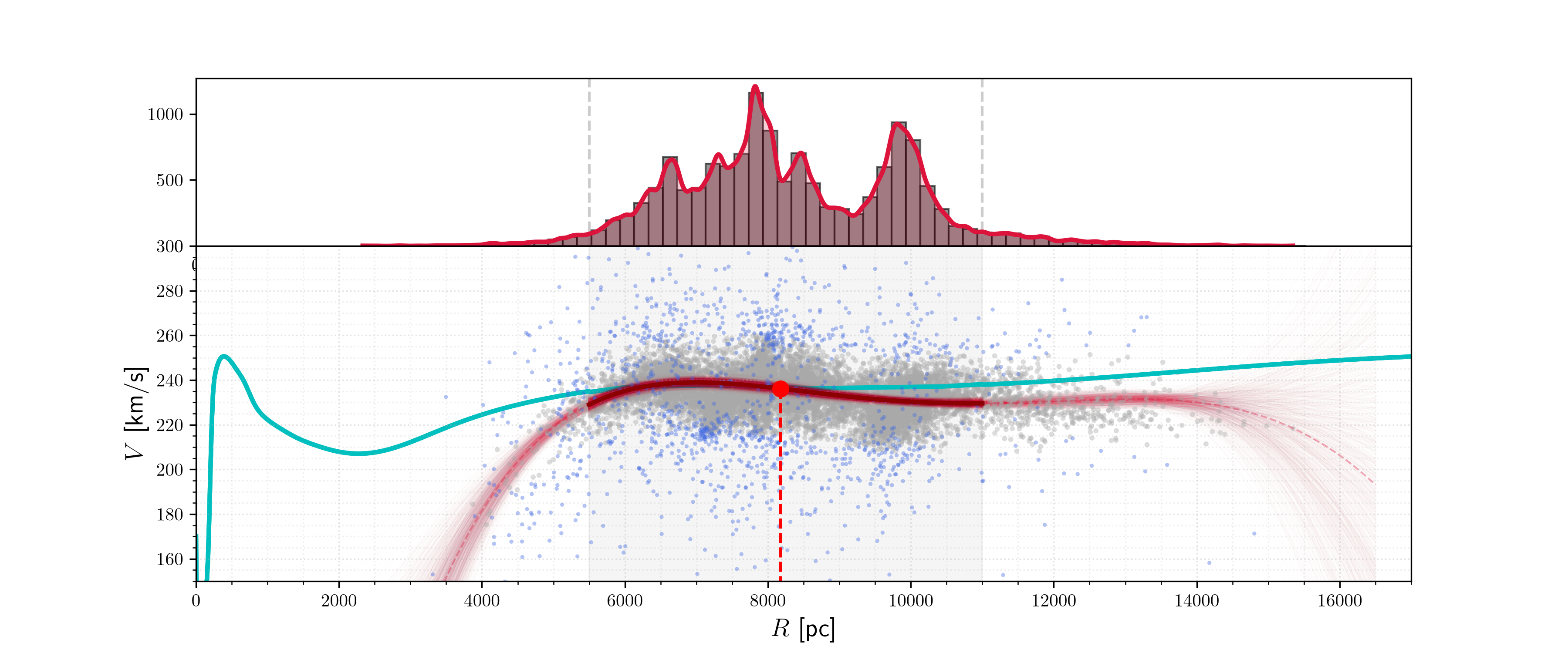}
\includegraphics[width=\textwidth]{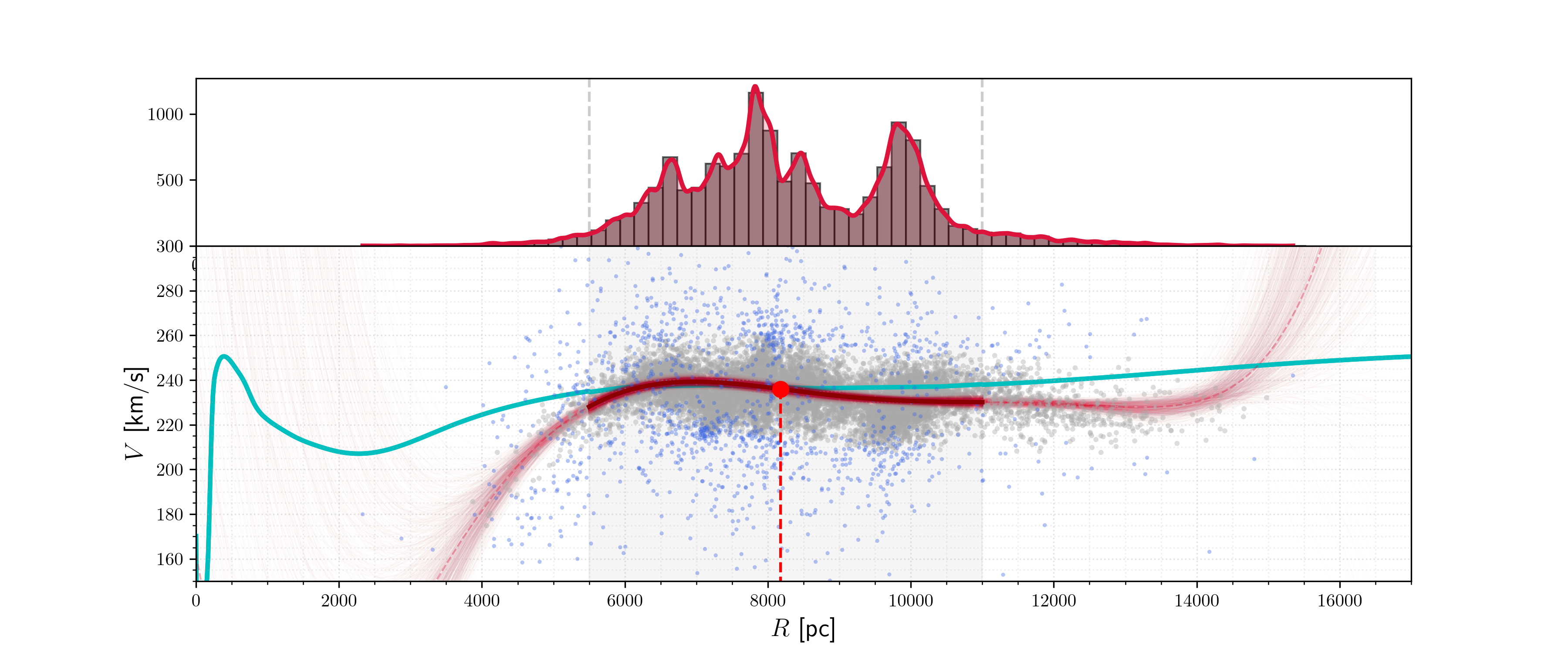}
\caption{Galactic rotation curve derived from the ALS III OB sample using polynomials of degree $3$ (top), $5$ (middle) and $6$ (bottom), to be compared with the adopted degree $4$ in Fig.~\ref{fig:rotcurve}. Symbols and shading are as in that figure. The curves are indistinguishable within the range where the sample provides statistical weight and diverge outside it.}
\label{fig:rotcurve_othergrades}
\end{figure*}

\end{appendix}

\end{document}